\documentclass[fleqn,usenatbib]{mnras}

\usepackage{newtxtext,newtxmath}

\usepackage[T1]{fontenc}

\DeclareRobustCommand{\VAN}[3]{#2}
\let\VANthebibliography\thebibliography
\def\thebibliography{\DeclareRobustCommand{\VAN}[3]{##3}\VANthebibliography}

\usepackage{graphicx}	
\usepackage{amsmath}	

\title[Outlook on Magnetohydrodynamical Turbulence]{Outlook on Magnetohydrodynamical Turbulence and its Astrophysical Implications}

\author[Popova \& Lazarian]{
Elena Popova$^{1}$\thanks{E-mail: elena.popova@ubo.cl},
Alexandre Lazarian$^{1,2}$\thanks{E-mail: alazarian@facstaff.wisc.edu}
\\
$^{1}$Centro de Investigación en Astronomía, Universidad Bernardo O’Higgins, Santiago, General Gana 1760, 8370993,
Chile\\
$^{2}$Department of Astronomy, University of Wisconsin-Madison, Madison, WI, 53706, USA\\
}

\date{Accepted XXX. Received YYY; in original form ZZZ}

\pubyear{2022}

\begin{document}
\label{firstpage}
\pagerange{\pageref{firstpage}--\pageref{lastpage}}
\maketitle

\begin{abstract}
Magnetohydrodynamical (MHD) turbulence is ubiquitous in magnetized astrophysical plasmas, and it radically changes a great variety of astrophysical processes. In this review, we give the concept of MHD turbulence and explain the origin of its scaling. We consider the implications of MHD turbulence to various problems: dynamo in different types of stars, flare activity, solar and stellar wind from different stars, propagation of cosmic rays, and star formation. We also discuss how the properties of MHD turbulence provide a new way of tracing magnetic fields in interstellar and intracluster media.

\end{abstract}


\section{Introduction}

Turbulence is ubiquitous astrophysics. The evidence of turbulence includes  a Kolmogorov spectrum of electron
density fluctuations \cite[see][]{Armstrong_etal:1995, ChepurnovLazarian:2010}.
 through the numerous measurements of the solar wind fluctuations \citep{Leamon_etal:1998} and
the non-thermal broadening of spectral lines as well as measures obtained by other
techniques \cite[see][]{Burkhart_etal:2010}. This is due to astrophysical plasmas  having very large
Reynolds numbers.  Plasma flows at these high Reynolds numbers are subject
to numerous linear and finite-amplitude instabilities that induce  turbulence.

Turbulence can be driven by an external energy source, such as
supernova explosions in the ISM \citep{NormanFerrara:1996, Ferriere:2001}, merger events and
active galactic nuclei outflows in the intercluster medium (ICM)
\citep{Subramanian_etal:2006, EnsslinVogt:2006, Chandran:2005}, and baroclinic
forcing behind shock waves in interstellar clouds. Turbulence can also be driven by a rich array of
instabilities, such as magneto-rotational instability (MRI) in accretion disks
\citep{BalbusHawley:1998, JafariVishniac2018disks}, kink instability of twisted flux tubes in the solar
corona \citep{GalsgaardNordlund:1997, GerrardHood:2003}, etc.  

The properties of the media change dramatically in the presence of turbulence. In particular, transport processes are changed by turbulence. 

Astrophysical turbulence is magnetized and therefore the turbulence in the presence of magnetic field is most important for astrophysical applications. In this review we discuss the basic properties of MHD turbulence in \S 2, its relation to dynamo in \S 3, to magnetic reconnection and reconnection diffusion in \S 4. We discuss the turbulence in spiral galaxies and its effects on star formation in \S 5. Turbulence effects on cosmic ray physics are considered in \S 6, while a new way of magnetic field studies that employs MHD turbulence properties, i.e., the gradient technique (GT) is discussed briefly in \S 7.

\section{Basics of MHD turbulence theory}

\subsection{General considerations}
Modern understanding of MHD turbulence theory is described in the monograph by Beresnyak and Lazarian \cite{BL19}. Below we provide a brief sketch of a few fundamental ideas at the theory's foundations.  

First of all, MHD turbulence is anisotropic. This property  has been known for a while (see \cite{1996JGR...101.7619M}) but was associated with scale-independent anisotropy that was measured in numerical simulations.  In \cite{GS95}, the concept of scale-dependent anisotropy scaling was introduced. The compressible MHD turbulence  is based on the notion of a superposition of 3 cascades of fundamental modes, i.e., Alfv\'en, slow and fast. We use  term "mode" rather than "wave" as in strong turbulence, the Alfv\'en turbulent perturbations undergo non-linear damping/cascading over one period. This definitely not wave behavior. 

\begin{figure*}
	\centering
	\includegraphics[width=.80\linewidth]{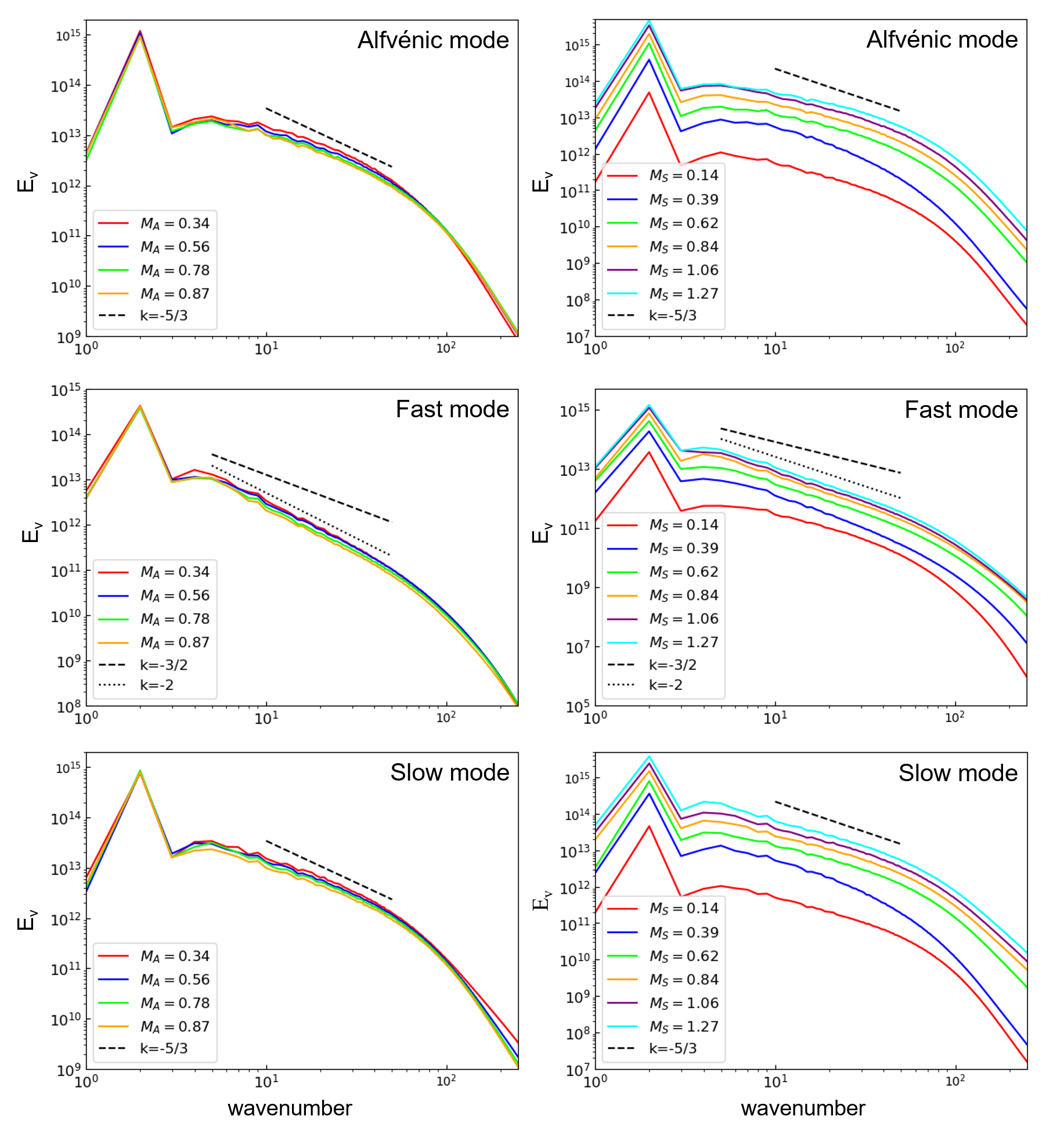}
 \caption{Left: The power spectrum of velocity fluctuations for Alfv\'enic (top), fast (central), and slow (bottom) modes. The sonic Mach number $M_S\approx0.6$. Right: The spectra of 3 modes for $M_A\approx0.5$. From Hu et al. (2022a). }
 \label{fig:2}
\end{figure*}

The Alfv\'en modes determine turbulence anisotropy. They slave the cascade of slow modes and impose their anisotropy on the slow modes \citep{GS95,LG01,CL02_PRL,CL03}. In non-relativistic turbulence fast modes follow their own cascade that depends marginally on the cascades of Alfven and slow modes \citep{CL02_PRL}. This cascade is similar to the acoustic one in high $\beta$ ($\beta$ is the ratio of the plasma to magnetic pressure) medium (\cite{GS95}). In this regime, the fast modes are mostly compressions of media propagating with sound velocity $c_s$.  \cite{CL02_PRL} demonstrated that in the low-$\beta$ turbulent  the fast modes form a cascade similar to the acoustic one, even though the fluctuations are  magnetic compressions propagating with Alfven velocity $V_A$. In fact, numerical simulations in \cite{CL02_PRL,CL03} indicate that the cascade of the fast ways is very similar to the acoustic cascade for all $\beta$. Fig.~\ref{fig:2} illustrates the spectra MHD modes. We discuss their properties below.

\subsection{Alfv\'en and slow modes}

The properties of MHD turbulence change with the Alfv\'en Mach number $M_A=V_L/V_A$, where $V_L$ and $V_A$ are, respectively, the turbulent injection and Alfv\'en velocities. For $M_A\ll 1$, magnetic fields are weakly perturbed, while magnetic fields are highly chaotic for $M_A \gg 1$.

The original version in \cite{GS95} was formulated by \cite{GS95} in Fourier space in the system of reference of the mean magnetic field.  In fact, the formulated scalings {\it are not valid in the system of the mean magnetic field}. 
The correct choice of the system of reference follows from the theory of turbulent reconnection \cite{LV99}. Turbulent reconnection is part and parcel of MHD turbulence (see \cite{Eyink_etal:2011}). \cite{LV99} theory demonstrates that in trans-Alvenic turbulence, the reconnection of eddies is equal to the eddy turnover time. Therefore Alfvenic turbulence can be treated as the collection of eddies whose axes of rotation are aligned with the magnetic field in their vicinity.  Turbulent reconnection enables the fluid motions that induce mixing perpendicular to the {\it local} magnetic field direction.

It is easy to see that the eddy's perpendicular magnetic field in the presence of turbulent reconnection experience minimal resistance. Indeed, the magnetic field bending by such eddies is minimal. Thus the energy of turbulent driving is being channeled through such eddies.

An eddy with scale $l_\bot$ perpendicular to the local magnetic field induces an Alfv\'{e}nic perturbation of scale $l_\|$ that propagates along the magnetic field with speed $V_A$. As an eddy induces this perturbation with the turnover time $l_\perp/v_l$ this should be equal to the timescale of the corresponding Alfvenic perturbation $l_\|/V_A$ induced by the eddy:
\begin{equation}
\frac{l_\|}{V_A}\approx \frac{l_\perp}{v_l},
\label{critbal}
\end{equation}
This relation constitutes the modern understanding of the critical balance for Alfv\'enic turbulence.

As a result, the \cite{GS95} original theory must be augmented by a concept of {\it local system of reference}. The vital significance of this system of reference for describing MHD turbulence is confirmed by numerical simulations \citep{CV00,MG01,CLV_incomp}. One should keep in mind that the direction of the local magnetic field at a given region can differ significantly from the global mean magnetic field direction. The latter results from large-scale averaging, and it does not generally coincide with the realization of the magnetic field at a given point. Disregarding the difference between the local and global systems of reference is the most common mistake in the literature dealing with MHD turbulence. The scale-dependent anisotropy laws formulated in \cite{GS95} are valid only in the local reference system!

Observational studies of the average magnetic field along the line of sight make it impossible to define the 3D local reference frame. Thus the observationally-measured anisotropy is not expected to be scale-dependent. In the reference system related to the mean field, i.e. global system of reference, the largest eddies dominate the measured turbulence anisotropy (see \cite{Cho_etal:2002}).  

While the local system of reference is not easy to define in measuring magnetic field from observations, this is a natural system of reference for astrophysical processes. For instance, the local direction of magnetic field in the Solar vicinity is very different from the average direction of magnetic field of the Milky way. Cosmic ray propagation in the Solar system neighborhood is determined by the local magnetic field rather than the Milky Way averaged one.

As we mentioned earlier, turbulent reconnection allows the turbulent cascade in the direction perpendicular to the local magnetic field. This cascade is not affected by the back-reaction of the magnetic field. Thus the Alfvenic cascade is  Kolmogorov-like. For trans-Alfv\'{e}nic turbulence with $V_L = V_A$ this entails:
\begin{equation}
v_l\approx V_A \left(\frac{l_\bot}{L}\right)^{1/3},
\label{GS95perp}
\end{equation}
where $v_l$ is the turbulent velocity corresponding to the perpendicular eddy $l_\perp$ scale.  

In the {\it local system of reference}, combining Eq. \eqref{GS95perp} and Eq. \eqref{critbal}, it is easy obtain the scale-dependent anisotropy of trans-Alfv\'{e}nic turbulence: 
\begin{equation}
l_{\|}\approx L \left(\frac{l_\perp}{L}\right)^{2/3}.
\label{ll}
\end{equation}
This testifies that smaller eddies are more elongated along the local magnetic field. 

If turbulence is injected with $M_A>1$, the magnetic field is weak at injection scale $L$. Therefore, the super-Alfv\'{e}nic turbulence at large scales evolves along an isotropic Kolmogorov energy spectrum. However, as turbulent velocity decrease at smaller scales, i.e. $v_l^2\sim V_L (l/L)^{2/3}$, the effect of the magnetic field becomes important. At the scale
\citep{Lazarian06}
\begin{equation}
l_A=L M_A^{-3},
\label{eq:A12}
\end{equation}
$v_l$ becomes equal to $V_A$, and the turbulence transfers to the MHD regime. In fact, at scales smaller than $l_A$ turbulence be described by trans-Alfv\'enic scaling, provided that $L$ in Eqs. \eqref{GS95perp} and \eqref{ll}
is replaced by $l_A$. 

The analysis of literature shows that the researchers frequently miss that the \cite{GS95} scalings are not valid for  sub-Alfv\'{e}nic turbulence with $M_A<1$.
It was demonstrated in \cite{LV99} that in the vicinity of the injection scale, $L$ the sub-Alfv\'enic turbulence evolves  along a different type of cascade. This is regime is termed {\it weak} regime of Alfvenic turbulence. In this regime, the parallel scale of wave packets remains equal to the injection scale, i.e., $l_\|=L=const$, and the Alfven perturbations interact multiple times to get cascaded. 

For weak turbulence the scaling obtained in \cite{LV99} for the weak turbulence\footnote{Weak turbulence is weak in terms of the non-linear interactions.} under the assumption of the isotropic turbulence driving at $L$ is
\begin{equation}
v_l\approx V_L \left(\frac{l_\perp}{L}\right)^{1/2},
\end{equation}
which also corresponds to the subsequent detailed analytical study in \cite{Gal00}.

An important feather of weak Alfvenic cascade is that with the decrease of $l_\perp$, the intensity of interactions of Alfv\'{e}nic perturbations increases. This is counterintuitive as with the decrease of $l_\perp$  the turbulence decreases in its amplitude. Therefore, as shown in \cite{LV99}, at scale:
\begin{equation}
l_\text{tran}\approx L M_A^2,
\label{ltrans}
\end{equation} 
where $M_A<1$, the turbulence gets into the strong turbulence regime. For the sub-Alfv\'{e}nic MHD turbulence at $l<l_\text{tran}$:
\begin{equation}
v_l\approx V_L \left(\frac{l_\bot}{L}\right)^{1/3} M_A^{1/3},
\label{vAlf}
\end{equation}
and:
\begin{equation}
l_{\|}\approx L \left(\frac{l_\bot}{L}\right)^{2/3} M_A^{-4/3}.
\label{lpar}
\end{equation}
It is important to note that the relations above that were derived in \cite{LV99} differ from Eqs. \eqref{GS95perp} and \eqref{ll} for trans-Alfv\'{e}nic turbulence, i.e. for $M_A=1$, by the additional dependence on the Alfv\'en Mach number $M_A$.

\subsection{Fast modes}

The numerical decomposition of MHD turbulence into modes has demonstrated that the interaction between fast modes, on the one hand, and slow and Alfven, on the other hand, is relatively weak for low Alfven Mach number $M_A=V_l/V_A$ for non-relativistic MHD turbulence \cite{CL02_PRL}. The cascade of the fast modes, therefore, can be assumed independent of the cascades of the Alfven and slow modes. This cascade is similar to the acoustic one in high $\beta$ medium\footnote{We remind the reader that $\beta$ is the ratio of the plasma to magnetic pressure.} \cite{GS95}. This is because, in this regime, the fast modes are mostly compressions of plasmas that propagate with sound velocity $c_s$. It was also shown by \cite{CL02_PRL} that in the opposite limiting case of low $\beta$ plasmas, the fast modes are expected to form a cascade similar to the acoustic one, even though the fluctuations are compressions of the magnetic field that propagate with velocity $\sim V_A$. The numerical simulations in \cite{CL02_PRL, CL03} support the idea that the cascade of the fast ways is very similar to the acoustic cascade for all $\beta$.

For fast modes in high $\beta$ medium, the perturbations are similar to sonic waves. Thus their amplitude increases with the sonic Mach number $M_s$. For fast modes in low-$\beta$ medium, the increase of amplitude corresponds to the increase of $M_A$. The numerical results in \cite{CL03} are consistent to $E_f\sim k^{-3/2}$, while those in \cite{KowL10} are better fitted by $E_f\sim k^{-2}$. 

The difference can be accounted for by appealing to the analogy with the acoustic turbulent cascade, but it is not a settled issue.  For instance, new simulations shown in Fig.~\ref{fig:1} suggest that the spectrum of $k^{-2}$ for subsonic turbulence corresponding $M_s=0.6$. Whatever their exact spectral index the fast modes fluctuations are isotropic. 

\begin{figure*}[htb!]
	\centering
	\includegraphics[width=1.0\linewidth]{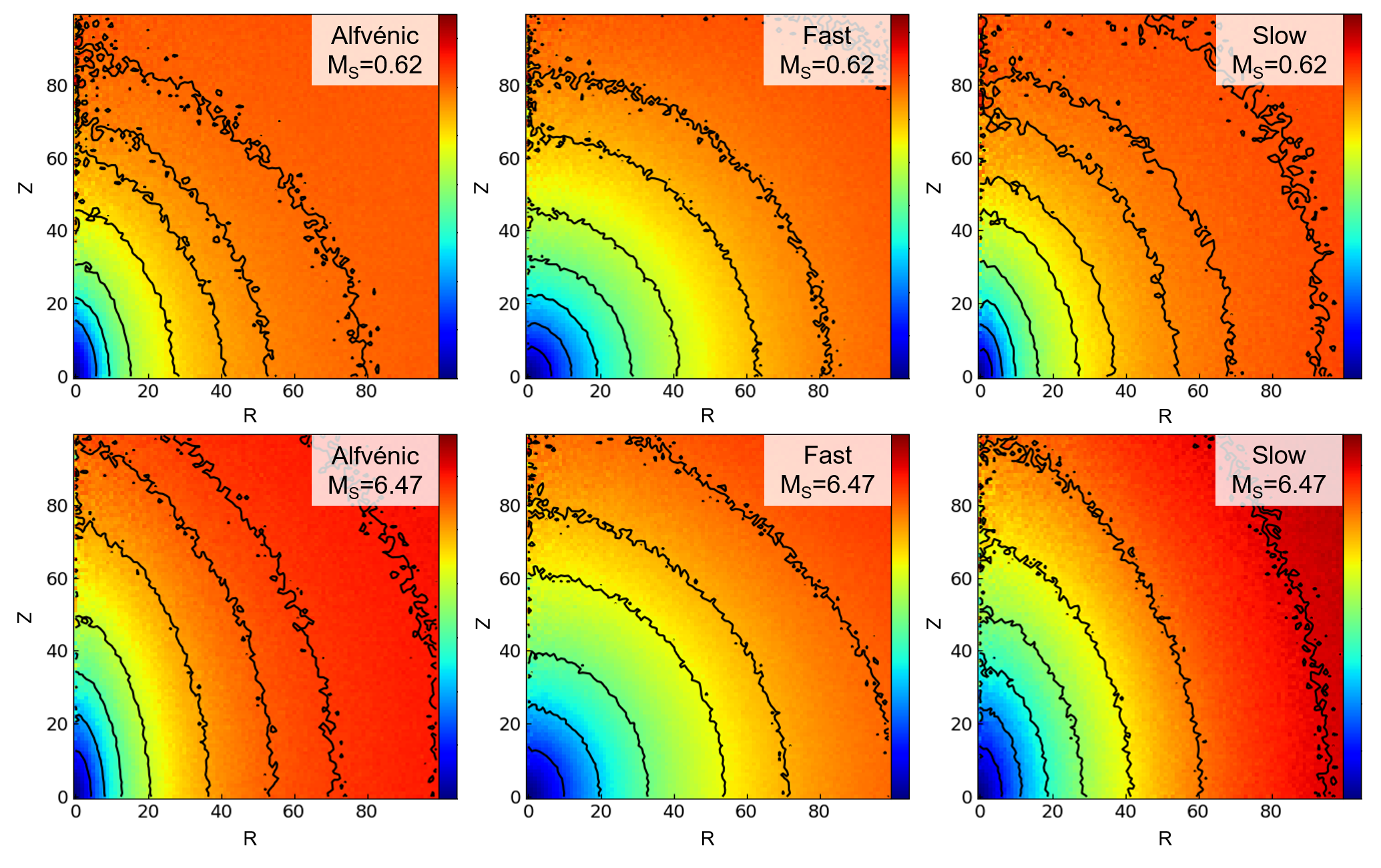}
	\caption{\label{fig:SF_map} Iso contours of equal correlation for structure functions measured in the local reference frame. The coordinate $Z$ is measured parallel to the local magnetic field while $R$ is measured perpendicular to the local magnetic field. The turbulence corresponds to $M_A\approx0.5$. From Hu et al. (2022a).}
\end{figure*}

Figure \ref{fig:SF_map} illustrates the anisotropies of 3 fundamental modes of MHD turbulence, i.e., Alfven, slow and fast. The contours of iso correlation are shown.

\section{MHD turbulence and dynamo}

MHD turbulence plays a key role in the turbulent dynamo. The dynamo process is a mechanism for generating a magnetic field in celestial bodies, in particular in the Sun and stars  \cite{Brandenburg2005}, \cite{Tobias2021}. The problem of explaining the occurrence of a magnetic field in celestial bodies began with the discovery of the terrestrial and solar magnetic fields. 

The first theoretical work on what constitutes the Earth's magnetic field, that is, what is the magnitude and direction of its intensity at each point on the earth's surface, belongs to the German mathematician Carl Gauss. In 1834, he gave a mathematical expression for the components of tension as a function of coordinates - the latitude and longitude of the observation site. Using this expression, it is possible to find for each point on the earth's surface the values of any of the components called the elements of the earth's magnetism. This and other works of Gauss became the foundation on which the edifice of the modern science of terrestrial magnetism is built \cite{Garland1979}. In particular, in 1839, he proved that the main part of the magnetic field comes out of the Earth, and the cause of small, short deviations in its values must be sought in the external environment \cite{Garland1979}.

The discovery of the solar magnetic field is associated with the observation of sunspots on its surface, which began to be conducted a very long time ago.

The first reports of sunspots date back to 800 BC. in China, sunspots are mentioned in the writings of Theophrastus of Athens (4th century BC), and the oldest known drawing of sunspots was made on December 8, 1128, by John of Worcester (published in The Chronicle of John of Worcester). In 1610, astronomers began using a telescope to observe the sun. Initial research focused on the nature of spots and their behavior. Despite the fact that the physical nature of the spots remained unclear until the 20th century, observations continued. In the 15th and 16th centuries, research was hindered by their small number, which is now regarded as a prolonged period of low SA, called the Maunder minimum. By the 19th century, there was already a sufficiently long series of observations of the number of sunspots to determine periodic cycles in the activity of the Sun. In 1845, Professors D. Henry and S. Alexander of Princeton University observed the Sun with a thermometer. They determined that the spots emit less energy than the surrounding areas of the Sun. Later, above-average radiation was determined in areas of the so-called solar plumes \cite{Arctowski1940}. 

For the first time, the magnetic field of the Sun was discovered and reliably measured in 1908 by J. Hale and just in one of the spots \cite{Hale2013}. Then the field strength turned out to be 2 kilogauss, which is 2-4 thousand times greater than the Earth's magnetic field (but almost 10 times less than the field of a modern magnetic resonance imaging apparatus, about 50 times less than the strongest fields created by man, and billions of times smaller than the fields of some neutron stars).

Now the observation of sunspots and the study of their magnetic fields is one of the daily tasks of modern heliophysics \cite{Mclntosh1972}-\cite{Hall1998}.

Now different agencies make monitoring of solar activity, for example, SDO: Solar Dynamics Observatory (https://sdo.gsfc.nasa.gov/). 

Note that other stars and their planets also have magnetic activity. There are many astronomical observations (KEPLER, TESS, and Evryscope-South Telescope), thanks to which large archives of observational data have been created. Stars have other modes of magnetic activity, for example, regular periods that are different from the solar, or without oscillations \cite{Baliunas1985}-\cite{Howard2020}.

Thus, observations of sunspots and then magnetic fields on the Sun, which have been carried out since the beginning of the 20th century, have shown that the intensity of magnetic fields varies, and these changes are cyclical. At the beginning of the 11-year solar cycle, the large-scale solar magnetic field is directed predominantly along the meridians (it is commonly said that it is "poloidal") and has an approximate dipole configuration. At the maximum of the cycle, it is replaced by a magnetic field of sunspots directed approximately along the parallels (the so-called "toroidal"), which at the end of the cycle is again replaced by a poloidal one - while its direction is opposite to that observed 11 years ago ("Hale's law").

The solar dynamo model is intended to explain the mentioned observed features. Since the conductivity of the solar plasma is quite high, magnetohydrodynamics describes the magnetic fields in the sun's convective zone. Due to the fact that the equatorial regions of the Sun rotate faster than the polar ones (this feature is called “rotation differential”), the initially poloidal field, being carried away by the rotating plasma, should stretch along the parallels, thereby acquiring a toroidal component. However, to ensure a closed, self-sustaining process, the toroidal field must somehow be transformed into a poloidal one. For some time, it was not clear how this happened. Moreover, Cowling's theorem explicitly forbade a stationary axisymmetric dynamo. In 1955, the American astrophysicist Eugene Parker, in his classic work \cite{Parker1955}, showed that the rising volumes of solar plasma must rotate due to the Coriolis forces, and the toroidal magnetic fields entrained by them can be transformed into poloidal ones (the so-called "alpha effect"). Thus, a model of a self-sustaining solar dynamo was constructed.

Currently, numerous solar dynamo models that are more complex than Parker's have been proposed, but, for the most part, go back to the latter. In particular, it is assumed that the generation of magnetic fields does not occur in the entire convective zone of the Sun, as previously thought, but in the so-called "tachocline" - a relatively narrow region near the boundary of the convective and radiant zones of the Sun, at a depth of about 200,000 kilometers under the solar photosphere, where the rotation speed changes sharply. The magnetic field created in this region rises to the surface of the Sun due to magnetic buoyancy.

The main dynamo models and their development is presented in the following works \cite{Moffatt1978}-\cite{Rincon2019}.

Dynamo problems deal with the physical description of the process of generating a magnetic field by a conducting fluid. Field generation is based on turbulence, and turbulent dynamo models are divided into "large scale/medium dynamo" and "small scale/fluctuating dynamo." In the first group, magnetic fields are amplified on scales larger than the outer scale of turbulence in the seconds on smaller scales \cite{Beresnyak2019}. In the second group, the small-scale turbulent dynamo is responsible for amplifying magnetic fields on scales smaller than the driving scale of turbulence in diverse astrophysical media \cite{Xu2021}.


In a turbulent dynamo, the amplification of magnetic fields occurs due to the turbulent stretching of magnetic fields (due to turbulent shear). 

The earliest work on the turbulent dynamo theory 
was done in \cite{Kazantsev1968}, \cite{Kraichnan1967} 
for the kinematic regime of the turbulent dynamo with a negligible back reaction of magnetic fields. In \cite{Kazantsev1968}, a simple model of the turbulent motion of a conducting liquid fluid was considered. The flow velocity has a Gaussian distribution function and the time for the establishment of diffusion of the liquid particles is zero. In this case, an exact solution of the problem of amplification of a spontaneous magnetic field can be derived. The instability criterion and magnetic field increment are obtained. In\cite{Kraichnan1967} the evolution of a weak, random initial magnetic field in a highly conducting, isotropically turbulent fluid is discussed with the aid of 
the exact expression for the initial growth of the magnetic energy spectrum. The possibilities of eventual growth and eventual decay are both admitted. For each, the shape of the magnetic‐energy spectrum in the case $\lambda>>\nu$ ($\lambda$ - magnetic diffusivity, $\nu$ - kinematic viscosity) is estimated by simple dynamical arguments. If there is growth, it is concluded that the magnetic spectrum below the Ohmic cut‐off eventually reaches equipartition with the kinetic‐energy spectrum with the principal exception that the spectrum of kinetic energy in the equipartition inertial range evolves to the form 
$k^{-3/2}$ and that equipartition is maintained, with the rapidly
falling spectrum, through part of the Ohmic dissipation range. The evolution of the magnetic spectrum in the weak‐field $\lambda>>\nu$ 
regime is also computed numerically with a simplified transfer approximation suggested by the Lagrangian‐history direct‐interaction equations. This calculation turns out to yield an eventual very weak exponential growth of magnetic energy.

In the case when the magnetic back reaction becomes significant, we have to consider the nonlinear turbulent dynamo. Nonlinear turbulent dynamo is characterized by the energy equipartition between turbulence and magnetic fields within the inertial range of turbulence \cite{xu2016turbulent}.

In \cite{chertkov1999small}, kinematic dynamo theory is presented for turbulent conductive fluids. The authors described how inhomogeneous magnetic fluctuations are generated below the viscous scale of turbulence, where the spatial smoothness of the velocity permits a systematic analysis of the Lagrangian path dynamics. In \cite{chertkov1999small} it was found the magnetic field's moments and multipoint correlation functions analytically at small yet finite magnetic diffusivity. The authors showed that the field is concentrated in long narrow strips and described anomalous scalings and angular singularities of the multipoint correlation functions, which are manifestations of the field's intermittency. The growth rate of the magnetic field in a typical realization is found to be half the difference of two Lyapunov exponents of the same sign.

In \cite{Schekochihin2004} authors showed results of an extensive numerical study of the small-scale turbulent dynamo. The primary focus is on the case of large magnetic Prandtl numbers $P_m$, which is relevant for hot low-density astrophysical plasmas. A $P_m$ parameter scan is given for the model case of viscosity-dominated (low Reynolds number $Re$) turbulence. The authors concentrated on three topics: magnetic energy spectra and saturation levels, the structure of the magnetic field lines, and the intermittency of the field strength distribution. In \cite{Schekochihin2004} main results are as follows: (1) the folded structure of the field (direction reversals at the resistive scale, field lines curved at the scale of the flow) persists from the kinematic to the nonlinear regime; (2) the field distribution is self-similar and appears to be lognormal during the kinematic regime and exponential in the saturated state; and (3) the bulk of the magnetic energy is at the resistive scale in the kinematic regime and remains there after saturation, although the magnetic energy spectrum becomes much shallower. The authors proposed an analytical model of saturation based on the idea of partial two-dimensionalization of the velocity gradients with respect to the local direction of the magnetic folds. The model-predicted saturated spectra are in excellent agreement with the numerical results. Comparisons with large-$Re$, moderate-$P_m$ runs are carried out to confirm these results' relevance and test heuristic scenarios of dynamo saturation. New features at large $Re$ are the elongation of the folds in the nonlinear regime from the viscous scale to the box scale and the presence of an intermediate nonlinear stage of slower than exponential magnetic energy growth accompanied by an increase of the resistive scale and partial suppression of the kinetic energy spectrum in the inertial range. Numerical results for the saturated state do not support scale-by-scale equipartition between magnetic and kinetic energies, with a definite excess of magnetic energy at small scales. A physical picture of the saturated state is proposed.

In \cite{xu2016turbulent, Xu2019} it was found a striking similarity between the dependence of dynamo behavior on Prandtl number $P_m$ in a conducting fluid and $R$ (a function of ionization fraction) in the partially ionized gas. In a weakly ionized medium, the kinematic stage is largely extended, including exponential growth and a new regime of dynamo characterized by linear-in-time growth of magnetic field strength, and the resulting magnetic energy is much higher than the kinetic energy carried by viscous-scale eddies. Unlike the kinematic stage, the subsequent nonlinear stage is unaffected by microscopic diffusion processes. It has a universal linear-in-time growth of magnetic energy with the growth rate as a constant fraction $3/38$ of the turbulent energy transfer rate, which agrees well with earlier numerical results. Applying the analysis to the first stars and galaxies, S. Xu  \cite{Xu2019} found that the kinematic stage can generate a field strength only an order of magnitude smaller than the final saturation value. But the generation of large-scale magnetic fields can only be accounted for by the relatively inefficient nonlinear stage and requires a longer than free-fall time. It suggests that magnetic fields may not have played a dynamically important role during the formation of the first stars. 

\begin{figure}
\centering
\includegraphics[width=0.68\linewidth]{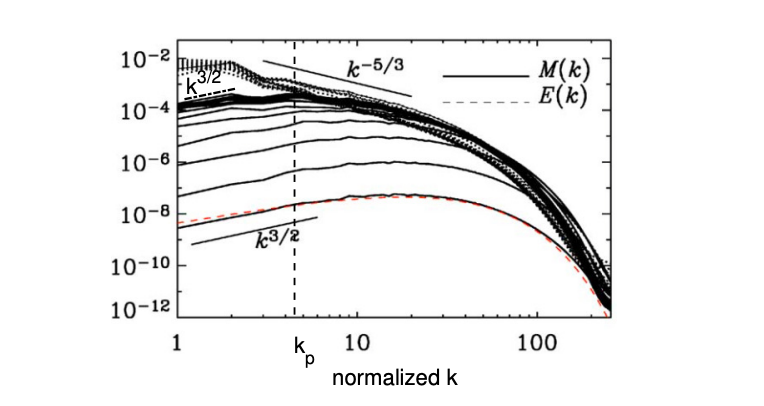}
\caption{Sketch of the magnetic (solid line) and turbulent kinetic (dashed line) energy spectra in the nonlinear stage of a turbulent dynamo for $P_m = 1$, dash-dotted lines indicate different spectral slopes and the vertical dashed line represents the equipartition scale at the end of their simulations. From Xu et al. (2019).
\label{image0}}
\end{figure}

Fig. \ref{image0} illustrates the magnetic energy spectrum in the nonlinear stage of the turbulent dynamo. At $P_m = 1$, it follows the Kazantsev ${k}^{3/2}$ profile on scales larger than $1/{k}_{p}$, while on smaller scales the transition to MHD turbulence occurs, and there is a ${k}^{-5/3}$ range for both the kinetic and magnetic energies.

\section{Magnetic reconnection and turbulence}

\subsection{Fast turbulent reconnection}
Turbulence accelerates many transport processes, e.g., those of heat and mass diffusion. The quantitative theory of turbulent reconnection was formulated in \cite{LV99}. The current state of this theory and its implications is summarized in \cite{lazarian20203d}.

\begin{figure}
\centering
\includegraphics[width=0.68\linewidth]{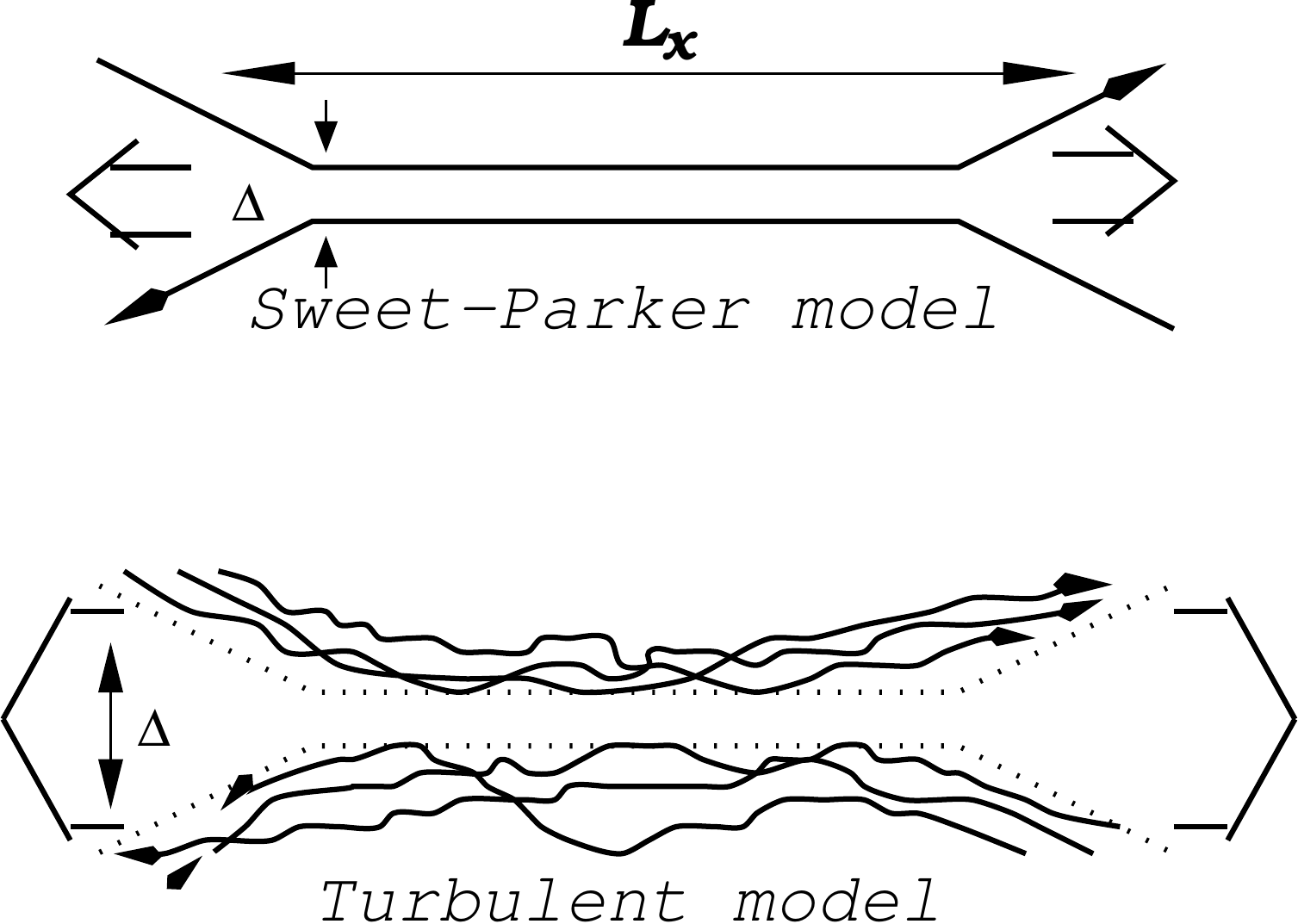}
\caption{{\it Upper plot}: Classical Sweet-Parker (SP) model of reconnection:  the thickness of the outflow $\Delta$ is limited by Ohmic diffusivity.  $L_x \gg \Delta$ makes the SP reconnection slow. 
{\it Lower plot}: From Lazarian and Vishniac (1999) model of reconnection includes turbulence in the SP setting. The outflow width $\Delta$ is determined by macroscopic field line wandering, and it can be $\sim L_x$ for trans-Alfvenic turbulence. From Lazarian and Vishniac (1999).}
\label{fig:recon}
\end{figure}

The \cite{LV99} model of turbulent reconnection is presented in Fig. \ref{fig:recon}. It is a natural generalization of the classical Swet-Parker model of reconnection in the case of turbulence. Both fluxes share the magnetic field of the same direction perpendicular to the figure's plane. The value of this so-called "guide field" does not change the reconnection rate (see \cite{lazarian20203d}). This magnetic field component is being ejected from the reconnection region with plasmas/fluid. 

\begin{figure}
\centering
\includegraphics[width=0.68\linewidth]{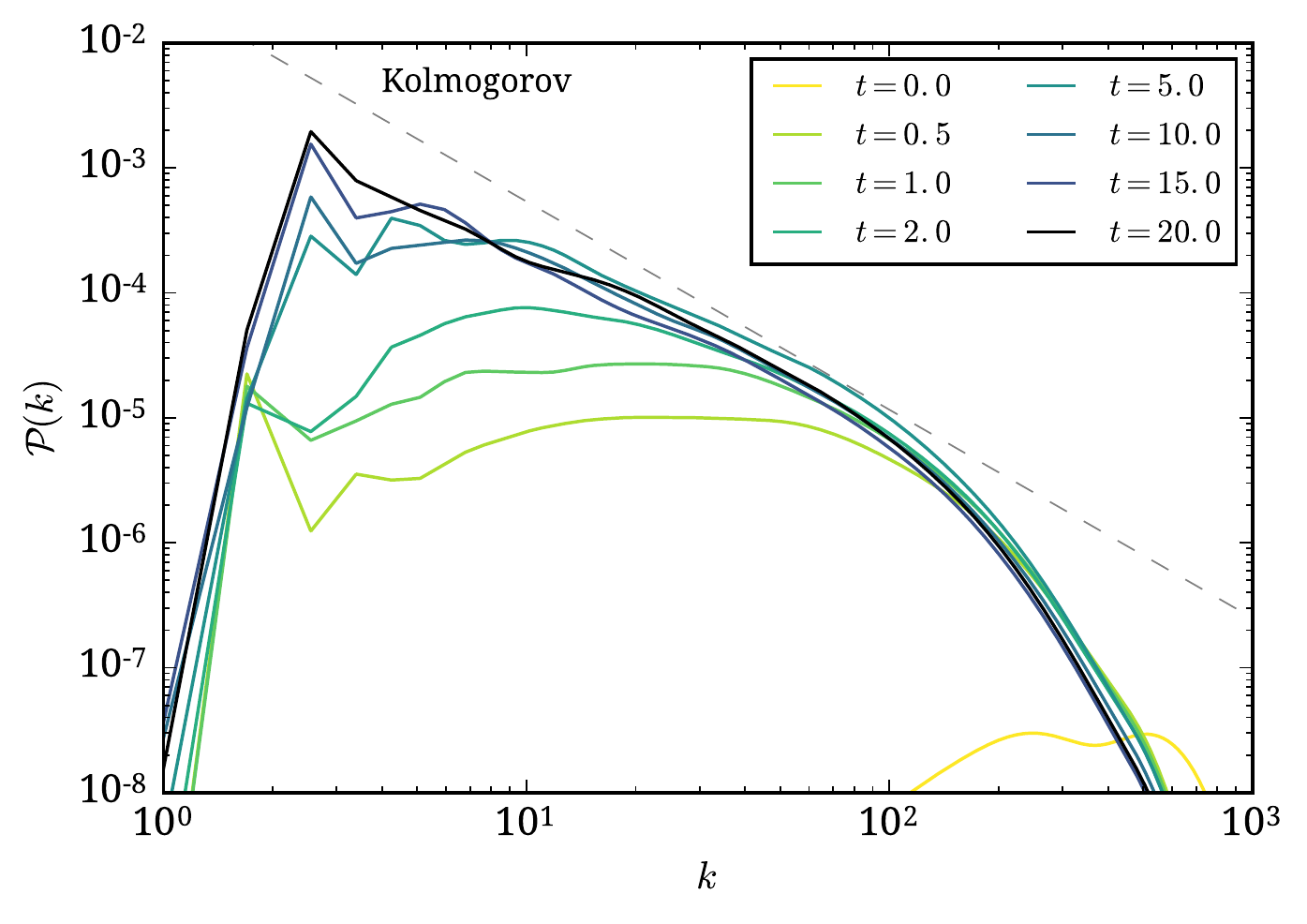}
\caption{Evolution of turbulent velocity power spectra of the self-induced reconnection. From Kowal et al. (2017).
\label{fig:kowal:spectra}}
\end{figure}

Unlike other models of fast magnetic reconnection, the \cite{LV99} model does not appeal to plasma effects but accounts for the magnetic field wondering induced by Alfvenic turbulence. Employing the Alfvenic mode scaling that we presented in \S 4 \cite{LV99} obtained the expression for the reconnection rate in the turbulence with injection at the scale $L$ and the opposite magnetic field in contact over scale $L_x$:
\begin{equation}
V_{rec} \approx V_A\min\left[\left(\frac{L_x}{L}\right)^{1/2},
\left(\frac{L}{L_x}\right)^{1/2}\right]
M_A^2.
\label{eq:lim2a}
\end{equation}
Note, that $V_A M_A^2$ is proportional to the turbulent eddy speed. As we discussed earlier, the obtained  reconnection
rate vary depending on Alfven Mach number $M_A$ and for $M_A\sim 1$ can represent a large fraction of the Alfv\'{e}n speed

It is clear from Eq. (\ref{eq:lim2a}) that the turbulent reconnection rate can be both slow and fast, depending on the system's turbulence level. This allows the reconnection model to explain various energetic phenomena that are impossible for the given prescribed reconnection rate. For instance, to explain the explosive release of magnetic energy in solar flares and gamma-ray bursts (see \cite{lazarian2019gamma}), it may be necessary to have periods of slow reconnection to accumulate the flux and periods of fast reconnection during which the energy is being released. The \cite{LV99} model was tested numerically (\cite{Kowal_etal:2009}, \cite{Kowal_etal:2012a}) and compared with observations (\cite{Sych_etal:2015}, \cite{ChittaLazarian:2019}). 

It is important that the outflow from the reconnection zone can induce turbulence, making turbulent reconnection self-induced. This process was sucessfully tested in a number of numerical studies (\cite{Kowal_etal:2017}, \cite{kowal2020kelvin}, \cite{Beresnyak:2017}, \cite{Oishi_etal:2015}). Figure \ref{fig:kowal:spectra} shows that the spectrum of turbulence induced by magnetic reconnection evolves towards the Kolmogorov-type cascade of Alfvenic turbulence.

Turbulent reconnection induces particle acceleration (\cite{lazarian2005production}). The predicted acceleration was successfully tested in \cite{Kowal_etal:2011}.

\subsection{Violation of Flux Freezing and Reconnection Diffusion}
The theorem by Alfven (1942) predicts that magnetic flux is frozen in within conductive fluid, i.e., the magnetic field and plasma move together. Turbulent reconnection violates the flux freezing condition. It was suggested in \cite{lazarian2005production} that in turbulent fluid magnetic field decouples from gas and can diffuse, solving, for instance, the problem of magnetic flux in star formation. More theoretical studies in \cite{Eyink_etal:2011} supported this idea. The flux freezing violation was demonstrated numerically in \cite{Eyink_etal:2013}.

\begin{figure}
\centering
\includegraphics[width=0.78\linewidth]{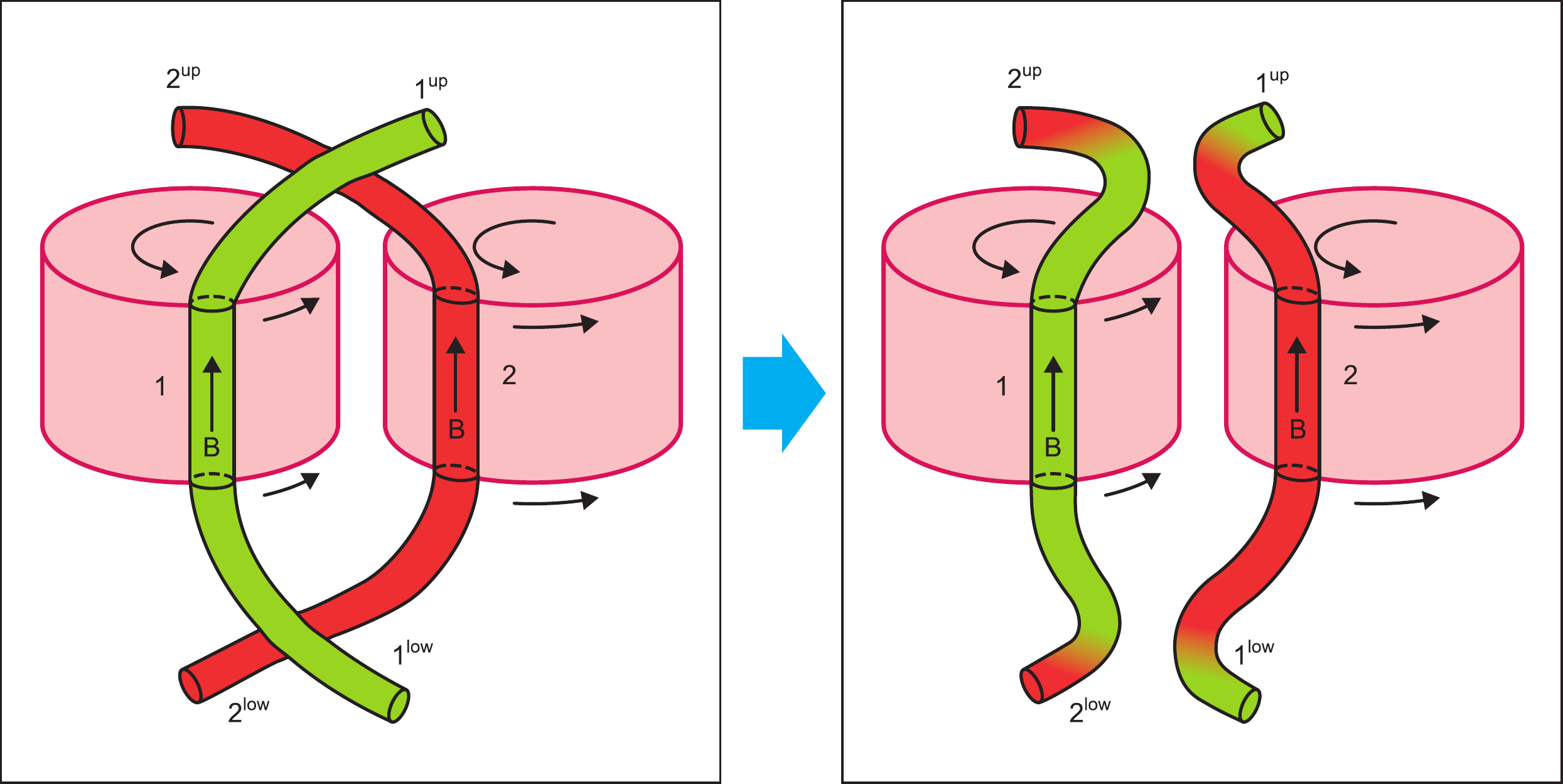}
\caption{Illustration of reconnection diffusion. Matter  and magnetic fields are exchanged  as two flux tubes of adjacent eddies interact.  From Lazarian et al. (2012), \copyright~AAS. Reproduced with permission.}
\label{fig30}
\end{figure}

Fast turbulent reconnection allows efficient exchange of magnetic field and plasmas between the adjacent turbulent eddies. This is illustrated in Figure \ref{fig30}, where the interaction of eddies of a given scale is shown. In reality, such interactions proceed at every scale of turbulent motions, which ensures efficient diffusion of magnetic fields.

\section{Turbulence in spiral galaxies}

\subsubsection{Properties of interstellar turbulence}

The interstellar media of spiral galaxies is turbulent. Figure \label{fig:1} presents the spectra of electron density fluctuation and velocity fluctuation that are observed in the Milky Way. Both power laws correspond to the Kolmogorov scaling, which has a natural explanation within MHD turbulence theory. 

\begin{figure*}
	\centering
	\includegraphics[width=0.45\linewidth]{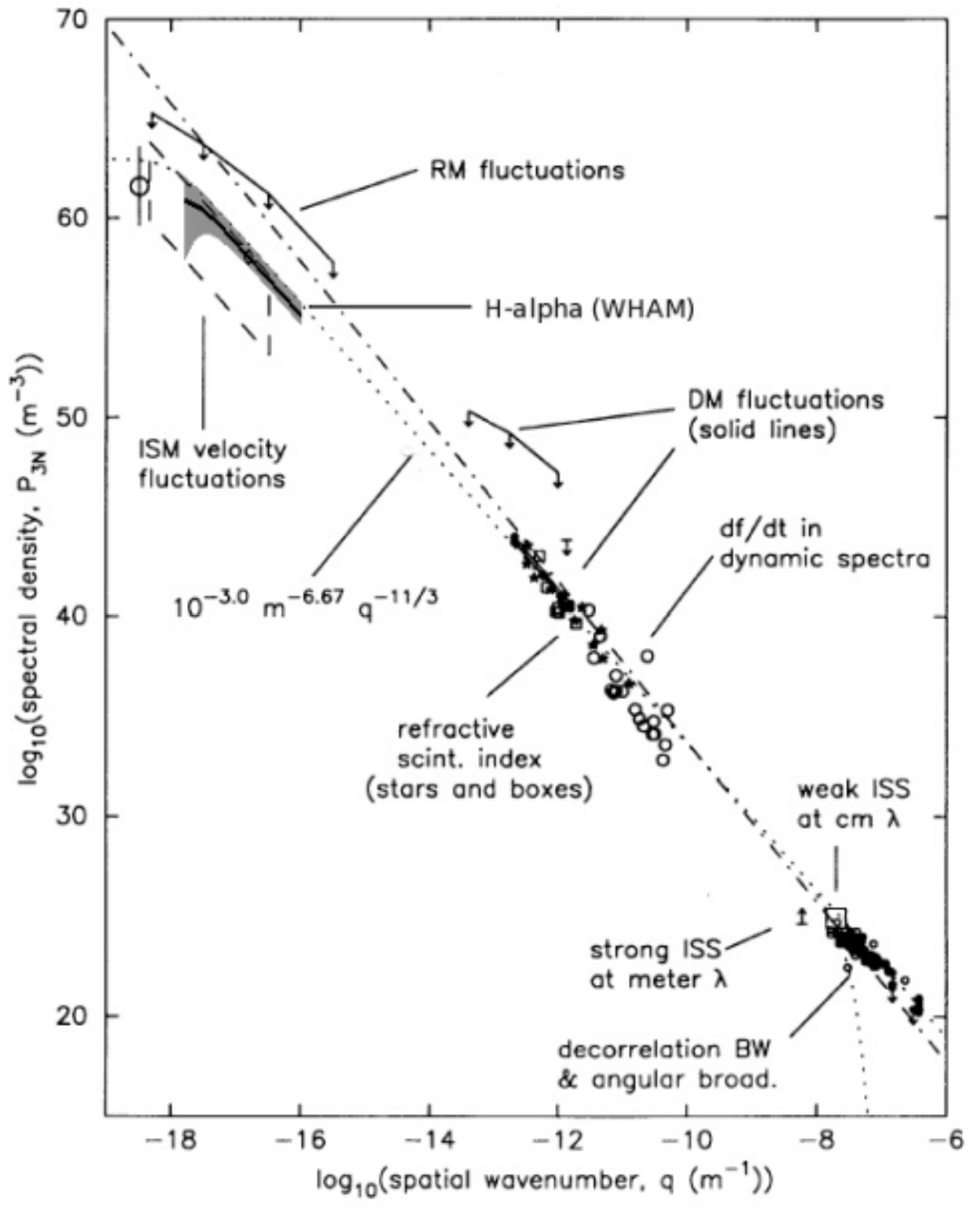}
 \includegraphics[width=0.45\linewidth]{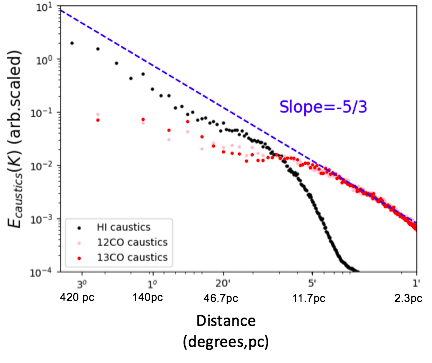}
\caption{\label{fig:1} . Left panel: "Extended Big Power Law" of galactic electron density fluctuations obtained combining the scattering measurements in Armstrong et al.(1995b) and $H_\alpha$ measurements from WHAM in Chepurnov and Lazarian (2010b). Right panel: Power low of velocity fluctuations in the direction to the Taurus molecular cloud. From Yuen et al. (2022).}
\end{figure*}

The fact that the velocity spectrum is Kolmogorov reflects the Kolmogorov scaling of perpendicular motions in Alfvenic turbulence. These motions dominate the observed cascade. As for the density fluctuations, the weakly compressible turbulence corresponding to the warm galactic gas passively reflect the statistics of velocity fluctuations. Note that for compressible parts of the media, steeper density spectra were observed (see \cite{xu2017scatter}).

\subsubsection{Effects on star formation}

Turbulence effects are essential for star formation. MHD turbulence has been used for classical theories of star formation as a supporting molecular cloud from the gravitational collapse. Such an approach required long-lived turbulent motions. Thus, the authors appealed to magnetic fields to mitigate turbulence decay. Alfvenic turbulence, as we discussed above, dissipates in one eddy turnover time. Thus, unless strong internal sources of turbulence driving exist, turbulence is hardly a means of supporting the typical molecular clouds from collapsing along magnetic field lines.

However, turbulence is good at doing a job that was not suspected. In the star formation theory, a big issue is the removal of the magnetic fields from collapsing clouds. In magnetically mediated star formation theory  (see \cite{MestelSpitzer:1956}, \cite{Mestel:1966}, \cite{Shu_etal:1987}, \cite{mouschovias1991magnetic}, \cite{nakano2002mechanism}, \cite{Shu_etal:2004}, \cite{Mouschovias_etal:2006})
magnetic fields counteract gravitational collapse. The magnetic flux freezing  is assumed to be frozen within the ionized component, while
the change in the flux-to-mass ratio is due to neutrals that flow pass ions and get concentrated towards the center of the gravitational potential. In a sense, ions act as guards that obey the magnetic field, while neutrals percolate through their ranks experiencing viscosity due to neutral-ion collisions. The latter process is termed in the star-formation literature "ambipolar diffusion". For decades ambipolar diffusion was assumed to be the necessary condition for star formation in the ISM. 

During ambipolar diffusion, the magnetic field resists the compression and leaves the gravitational potential, while neutrals get concentrated, forming the propostar  (e.g. \cite{Mestel:1965a}, \cite{Mestel:1965b}). The mediation of ambipolar diffusion was assumed to 
make star formation inefficient for magnetically dominated
(i.e., subcritical) clouds. The slow speed of ambipolar diffusion  entails low efficiency of star formation, which was interpreted as the strong observational support of the ambipolar diffusion paradigm (e.g., \cite{ZuckermanEvans:1974}).

This, however, does not solve all the
problems as for clouds dominated by gravity, i.e., supercritical clouds, the magnetic fields do not have time to leave the cloud through ambipolar diffusion. Therefore for such clouds, magnetic fields are expected to be dragged into the star, forming stars with
magnetizations far in excess of the observed ones (see \cite{Galli_etal:2006}, \cite{Johns-Krull:2007}).

The process of reconnection diffusion illustrated by Figure \ref{fig30} provides a viable explanation of the magnetic flux removal processes in the star formation process. The analytical prediction of the reconnection diffusion rate was confirmed numerically in \cite{santos2021diffusion}. This process induces fast magnetic flux removal independent of the media ionization degree. 

\begin{figure}
\centering
\includegraphics[width=0.78\linewidth, angle = -90]{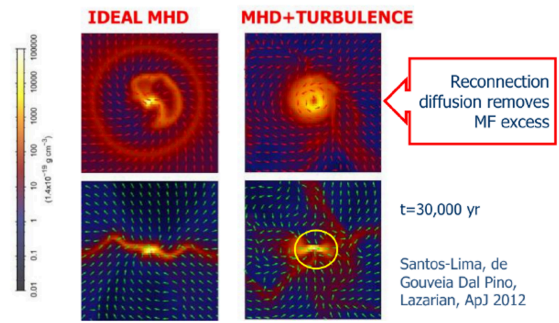}
\caption{{\it Left panels}. Evolution of disk without reconnection diffusion. The formed disk is much smaller compared to observed ones. {\it Right panels}. The disk produced via reconnection diffusion in 30 000 years. From Santos-Lima et al. (2010).}
\label{disk}
\end{figure}

The application of reconnection diffusion allows for solving the long-standing problem of magnetic flux removal from accretion disks, the so-called "magnetic braking catastrophe." The essence of the problem is that during circumstellar disk formation, the magnetic field of molecular clouds magnetic field is strong to transfer the matter momentum from the forming disk on a time scale shorter than the disk formation time. Figure \ref{disk} from numerical studies in \cite{Santos-Lima_etal:2010} show that the problem can be solved if reconnection diffusion is accounted for.      

\section{Turbulence and cosmic rays} 

Particles with energies ranging from MeV to PeV are usually termed cosmic rays (CRs). Their interaction with MHD turbulence controls the CR diffusion; their acceleration also depends on this interaction \cite{2011hea..book.....L}. The knowledge of CR propagation is vital for understanding the solar modulation of Galactic CRs, Fermi Bubble emission, and space weather forecasting \cite{Par65,Jo71,SinH01}. It is also essential for understanding of the origin of driving for galactic winds (e.g., \cite{Ipa75,Hol19}), and feedback heating in clusters of galaxies (e.g., \cite{Guo08,Brun14}). 

 CRs can interact with the pre-existing MHD fluctuations and the magnetic fluctuations created by them, e.g., by the perturbations created by the streaming instability (see \cite{Kulsrud_Pearce}). The suppression of streaming instability by MHD turbulence \cite{YL02,FG04,La16,2022ApJ...925...48X} can significantly modify the CR propagation \cite{2022FrP....10.2799L}. 
 
Earlier, the CRs interaction with magnetic turbulence was studied with ad hoc models adopted for MHD turbulence \cite{Jokipii1966,Kulsrud_Pearce,SchlickeiserMiller,Giacalone_Jok1999}. Those involved  the model of isotropic MHD turbulence (see \cite{Schlickeiser02}) as well as 2D + slab model for solar wind turbulence
\cite{Mat90}. Those, however, do not correspond to the understanding of the modern MHD turbulence presented in \S 2.

For CR propagation one, distinguishing the propagation perpendicular to the mean magnetic field and the propagation parallel to the mean magnetic field is useful. The perpendicular propagation is governed by magnetic field wondering described in \cite{LV99}. This induces processes of diffusion and superdiffusion, as illustrated in Table 1 from \cite{LazarianYan:2014}:
\begin{figure*}
	\centering
	\includegraphics[width=.70\linewidth, angle = -90]{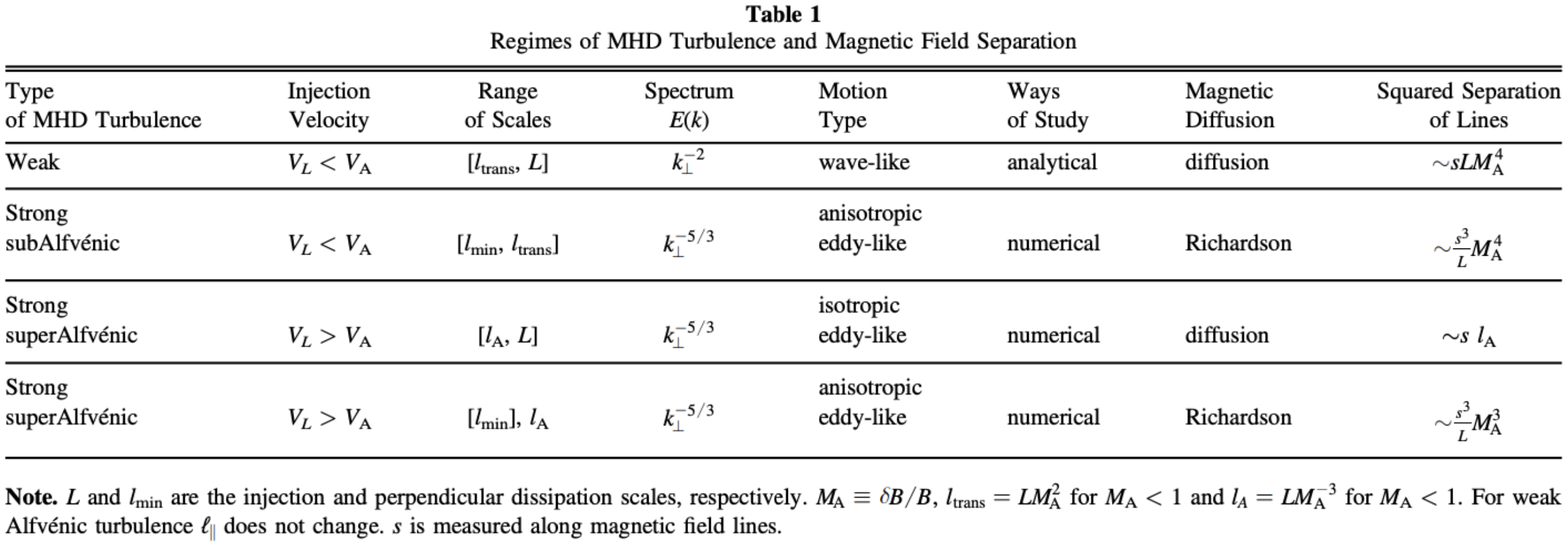}
	\caption{TABLE 1.}
\end{figure*}

\begin{figure}
	\centering
	\includegraphics[width=1.0\linewidth]{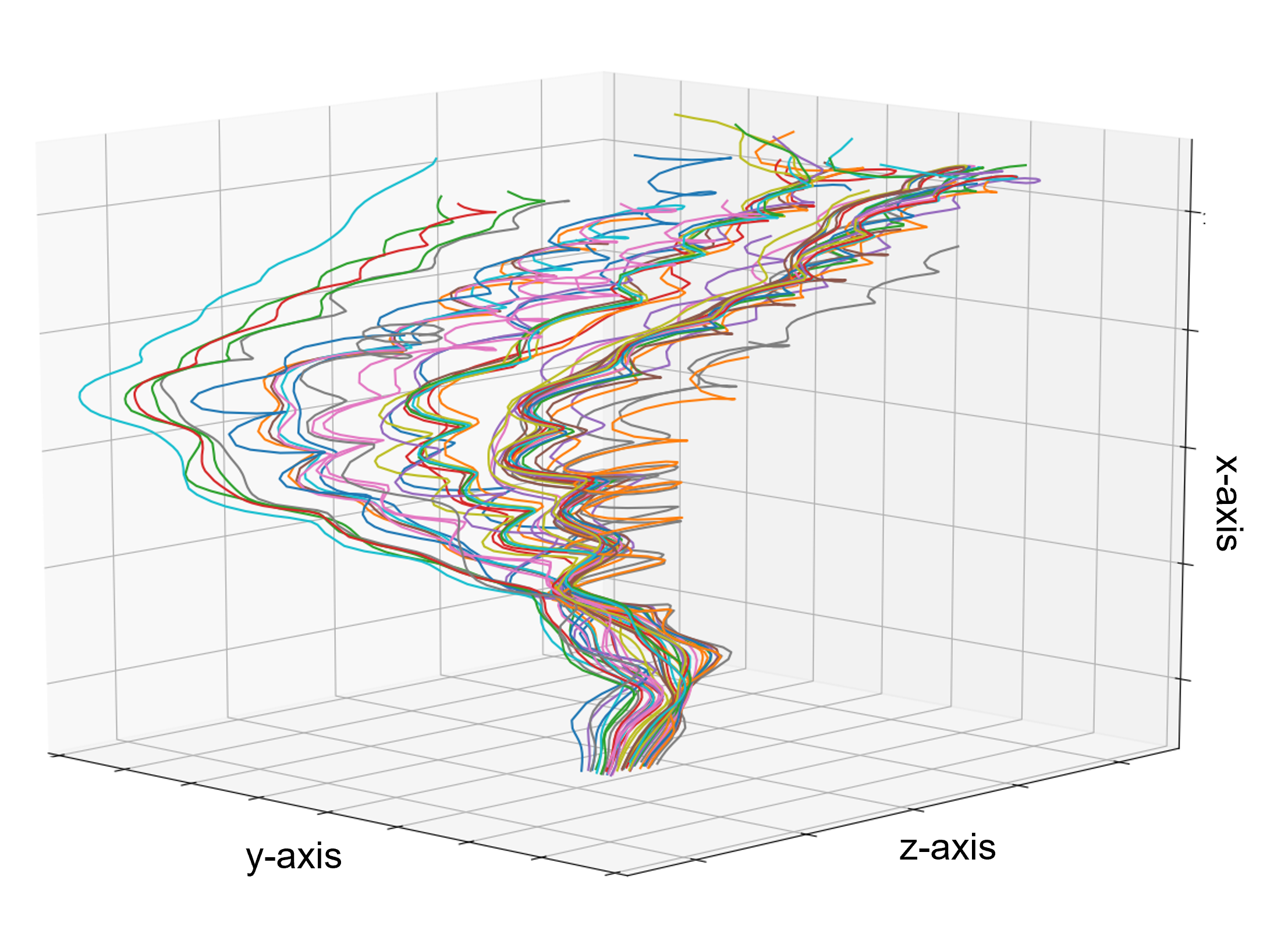}
	\caption{\label{fig:exmaple} CRs' superdiffusion: CRs' trajectories are shown with $M_S=0.62$ and $M_A=0.56$.  The initial spatial separation between CRs is one pixel and initial pitch angle is 0 degree. From Hu et al. (2022b).}
\end{figure}

The CR superdiffusion relates the transposition of the CR along the magnetic field with the transposition in the perpendicular direction. The accelerated diffusion of CRs in the perpendicular direction is illustrated by Figure \ref{fig:exmaple}. The superdiffusion acts on the scales less than the turbulence injection scale, radically changing the CR dynamics. 

\begin{figure}
\centering   
\includegraphics[width=0.45\linewidth]{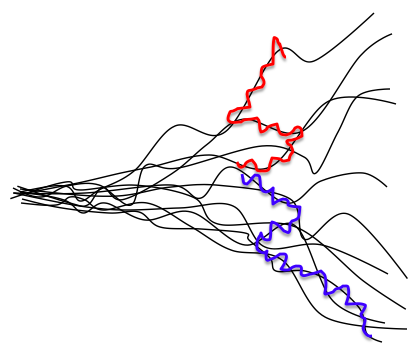}
 \includegraphics[width=0.45\linewidth]{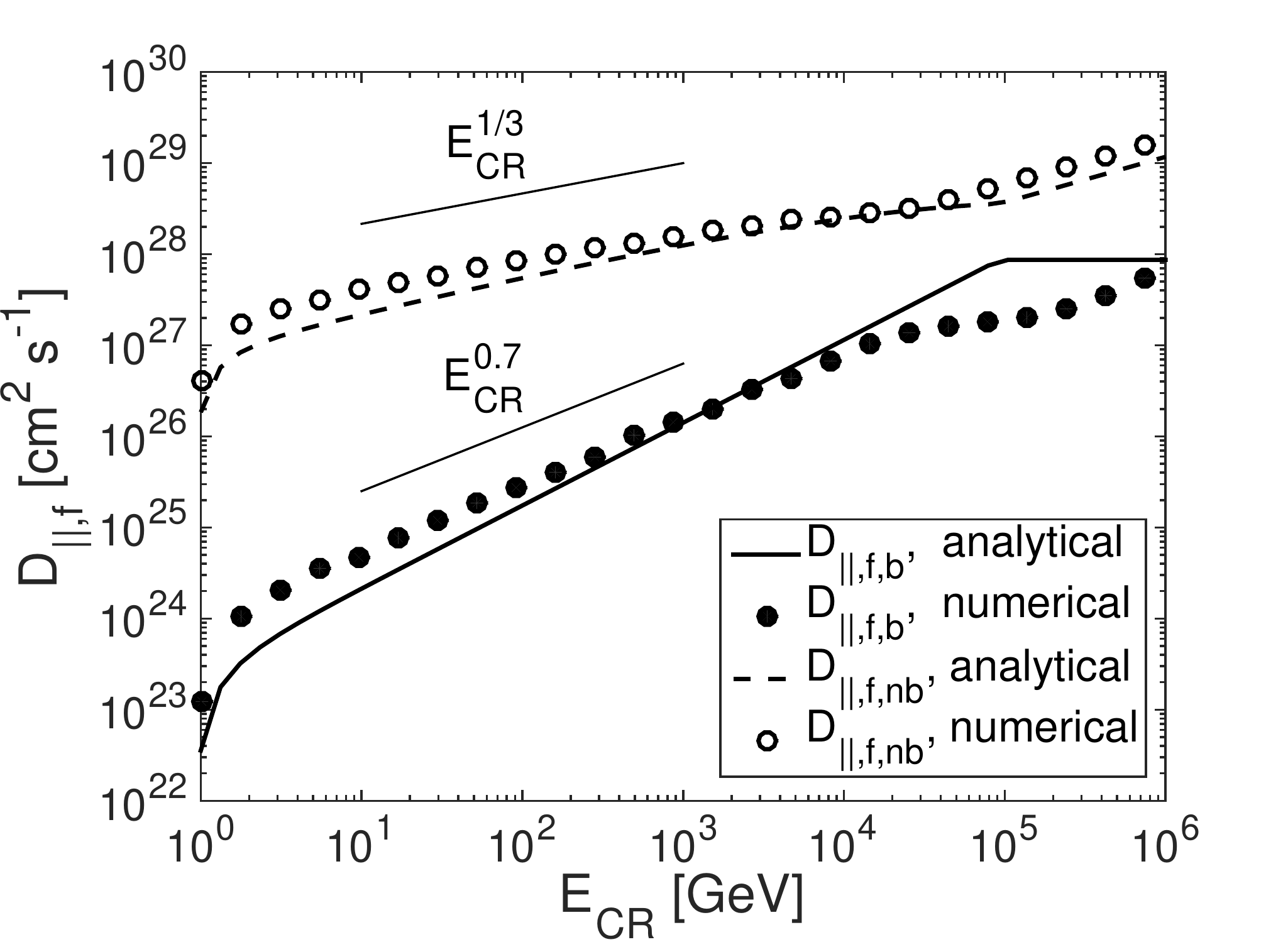}
\caption{{\bf Left Panel} Schematic of bouncing diffusion.  Trajectories of two particles with small initial separation are shown. {\bf Right panel} The comparison of the parallel diffusion coefficients induced by fast modes  $D_{\|,f,b}$ of bouncing CRs 
and $D_{\|,f,nb}$ of non-bouncing CRs. From Lazarian and Xu (2021).}
\label{fig: untrfast}
\end{figure}

In terms of parallel to magnetic field diffusion, pitch angle scattering and Transient Time Damping (TTD) are generally accepted processes \cite{Schlickeiser2002}. The first process arises from the resonant scattering of particles both compressible and incompressible fluctuations. In \cite{LY02} pitch angle scattering by fast modes was identified as the dominant process of scattering. The second is from particles surfing the compressible magnetic fluctuations which in many cases arise from particles surfing slow modes \cite{XLb18}. Understanding the CR perpendicular superdiffusion allowed \cite{LX21} to introduce a new process they termed {\it bouncing diffusion}. This type of diffusion arises from the simultaneous action of reflection from magnetic mirrors induced by slow and fast modes and the perpendicular superdiffusion. The bouncing diffusion acts on particles with small pitch angles this slows their diffusion. A comparison of the diffusion coefficients for bouncing and not bouncing particles is provided in Fig \ref{fig: untrfast}.

With fast modes providing efficient isotropization of CRs in terms of pitch angle $\mu$, \cite{LX21} evaluated the total distribution of scattering and bouncing particles as
\begin{equation}\label{eq: dtave}
   D_{\|,\text{tot}}\approx \mu_c^2 D_{\|,b} + (1-\mu_c^2) D_{\|,nb},
\end{equation}
where $D_{\|, scat}$ is the diffusion coefficient arising from scattering. The bouncing diffusion prevents the fast escape of particles with $\mu<\mu_{cr}$. Individual particles can generally exhibit periods of slow bouncing diffusion separated by periods of fast diffusion when they are in the scattering regime, i.e., Levi flights.

\section{Implication of turbulence: Gradient technique for studies of magnetic fields}

Studies of astrophysical magnetic fields  rely on the effects of the magnetic field on the media. Mapping plane-of-sky magnetic fields can be obtained with polarization from grains aligned with long axes perpendicular to the magnetic field \cite{andersson2015interstellar}, and synchrotron polarization \cite{beck2016magnetic}. 

\begin{figure}
	\centering
	\includegraphics[width=0.8\linewidth, angle = -90]{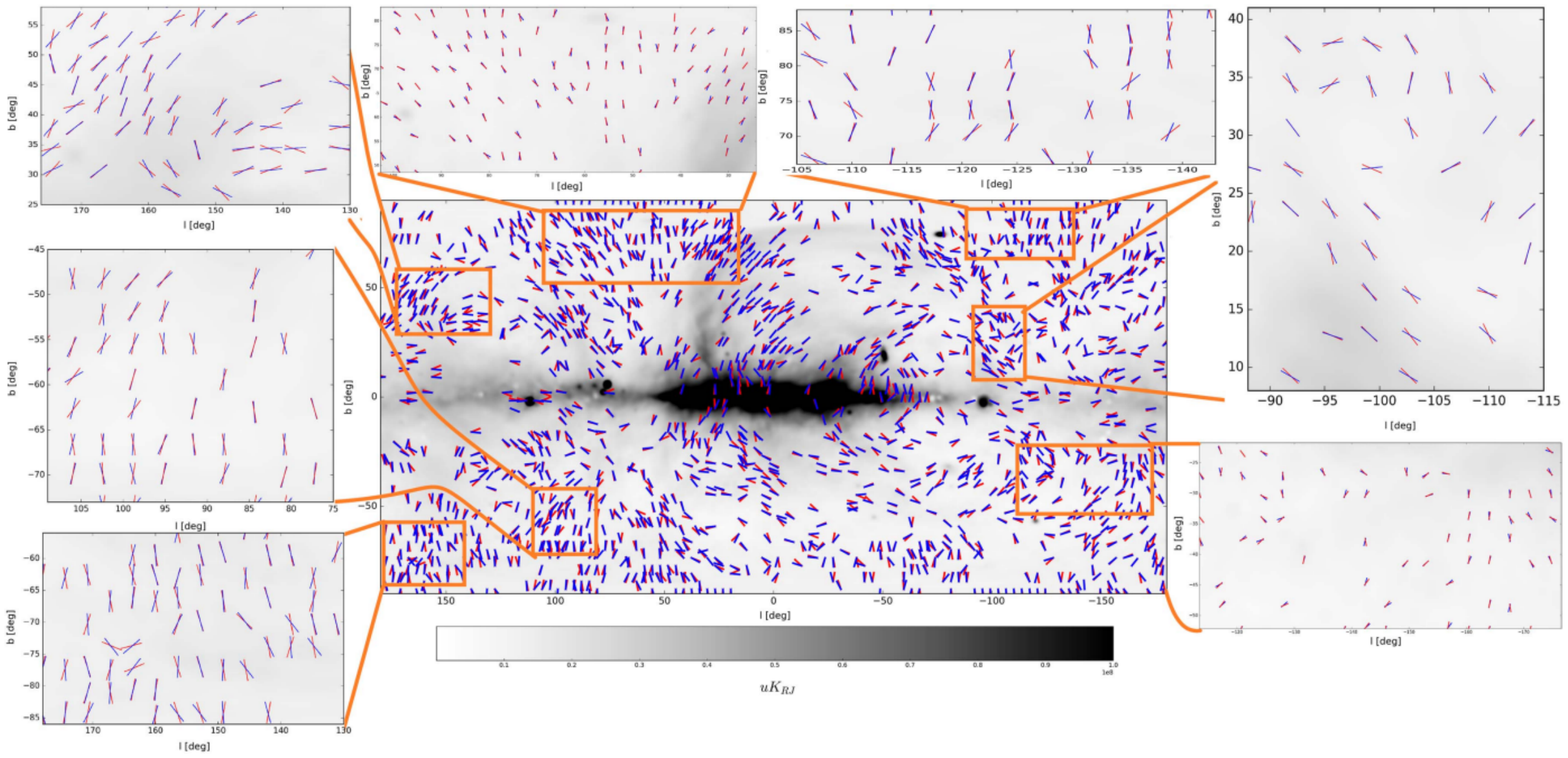}
	\caption{\label{fig:galaxy} Comparison of galactic magnetic fields obtained with polarization and SIGs.
 From Lazarian et al. (2017).}
\end{figure}

The above techniques employ polarization, and polarization measurements require significantly more effort than measurements of signal intensities. Therefore the Gradient Technique (GT) that allows mapping the Plane of Sky (POS) magnetic field without polarization measurements is opening a new avenue for magnetic field studies in diffuse media. The GT employs the properties of magnetic turbulence. The versions of the technique that do not require polarization measurements have been implemented as {\it Velocity Gradient Technique (VGT)} with subdivision of Velocity Centroid Gradients (VCGs) that employ Velocity Centroids (\cite{gonzalez2017velocity}, \cite{yuen2017tracing}), Velocity Channel Gradients (VChGs) \cite{LazarianYuen:2018a} that employ intensity fluctuations in thin channel maps, as well as {\it Synchrotron Intensity Gradients (SIGs} \cite{lazarian2017synchrotron} that employ synchrotron intensities. Note that GT can also employ polarization to get extra information about the magnetic field. For instance, as shown in \cite{LazarianYuen:2018b}, Synchrotron Polarization Gradients (SPGs) can use synchrotron polarization at different wavelengths to probe magnetic fields at different distances along the line of sight (see \cite{ho2019comparison}), while Faraday Gradients (FGs) can get the distribution of plane of sky direction of the magnetic field. However, we do not discuss polarization versions in the present paper. 

\begin{figure}
	\centering
	\includegraphics[width=0.8\linewidth, angle = -90 ]{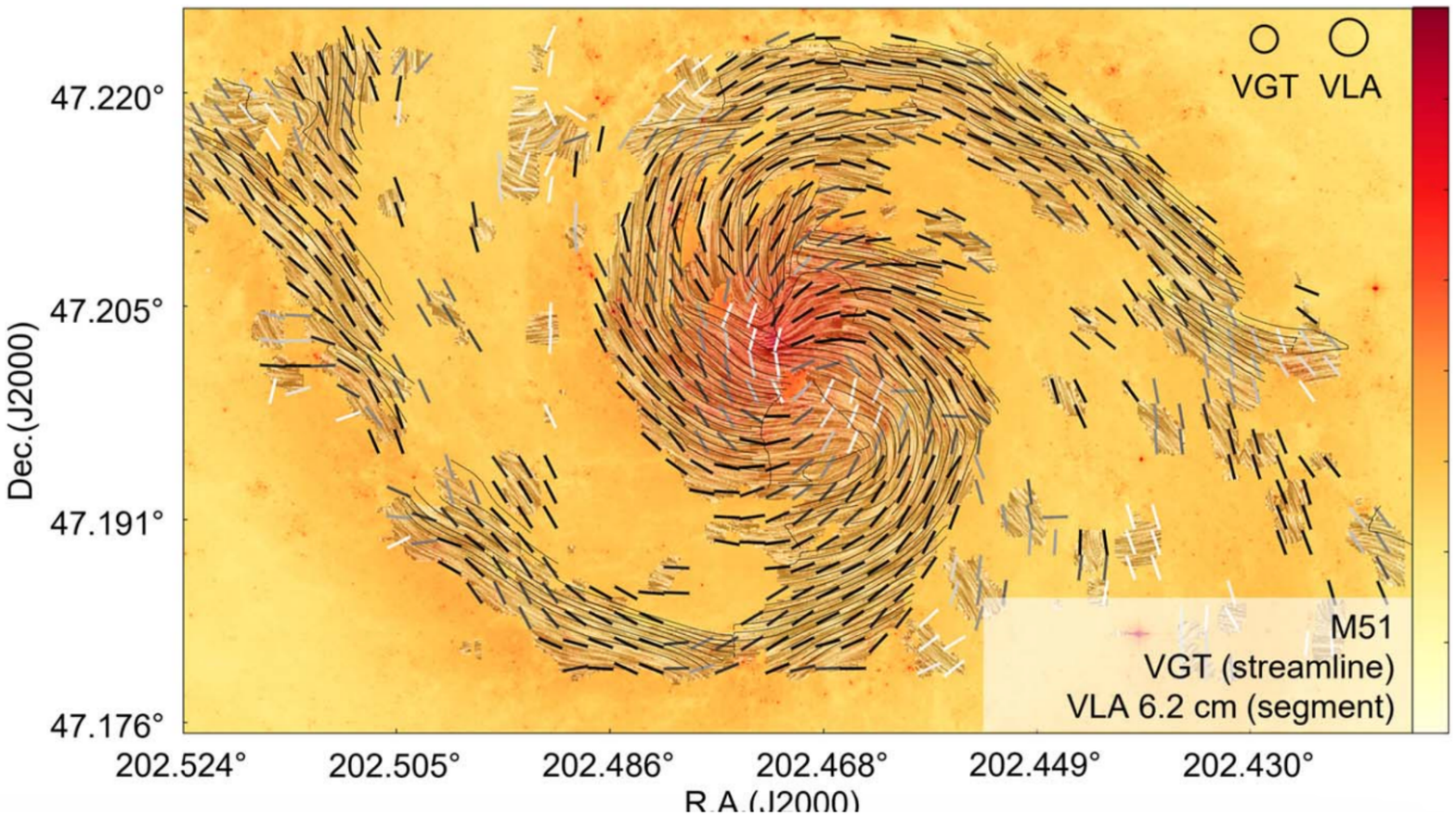}
	\caption{\label{fig:M51} Velocity gradients obtained from publicly available $^{12}CO$ maps from ALMA provide better resolution magnetic field maps compared to VLA.
 From Hu et al. 2022c.}
\end{figure}

Due to MHD turbulence's properties, as discussed in \S 3, the eddies are aligned with the magnetic field. The eddies' rotation along the magnetic field's local direction induces velocity and magnetic field gradients perpendicular to the magnetic field. In Kolmogorov-type turbulence of Alfvenic and slow modes, the gradients increase with the decrease of eddy scale as $v_l/l_\bot\sim l_\bot^{-2/3}$. Thus the smallest resolved eddies well aligned with the magnetic field dominate the gradients. Thus the gradients are perpendicular to the local direction of the magnetic field, revealing the magnetic field structure. 

Figure \ref{fig:galaxy} demonstrates the power of synchrotron gradients (SIGs). The maps of the magnetic field obtained with Planck synchrotron polarization are compared to those obtained with SIGs. In most cases, the difference in the obtained directions is negligible. A big advantage of SIGs is that they are not subject to Faraday rotation distortions. Those are especially harmful for low frequencies.

Figure \ref{fig:M51} shows the magnetic field structure of the active galaxy M51 mapped by velocity gradients (VGT) and synchrotron polarization. The structure of the galactic magnetic field is better resolved with gradients compared to VLA synchrotron polarization measurements. Comparing the magnetic field maps obtained with the VGT and dust polarization in \cite{hu2022role} reveals important details of the effects of the magnetic field on the central black hole accretion. 

\begin{figure}
	\centering
	\includegraphics[width=1\linewidth]{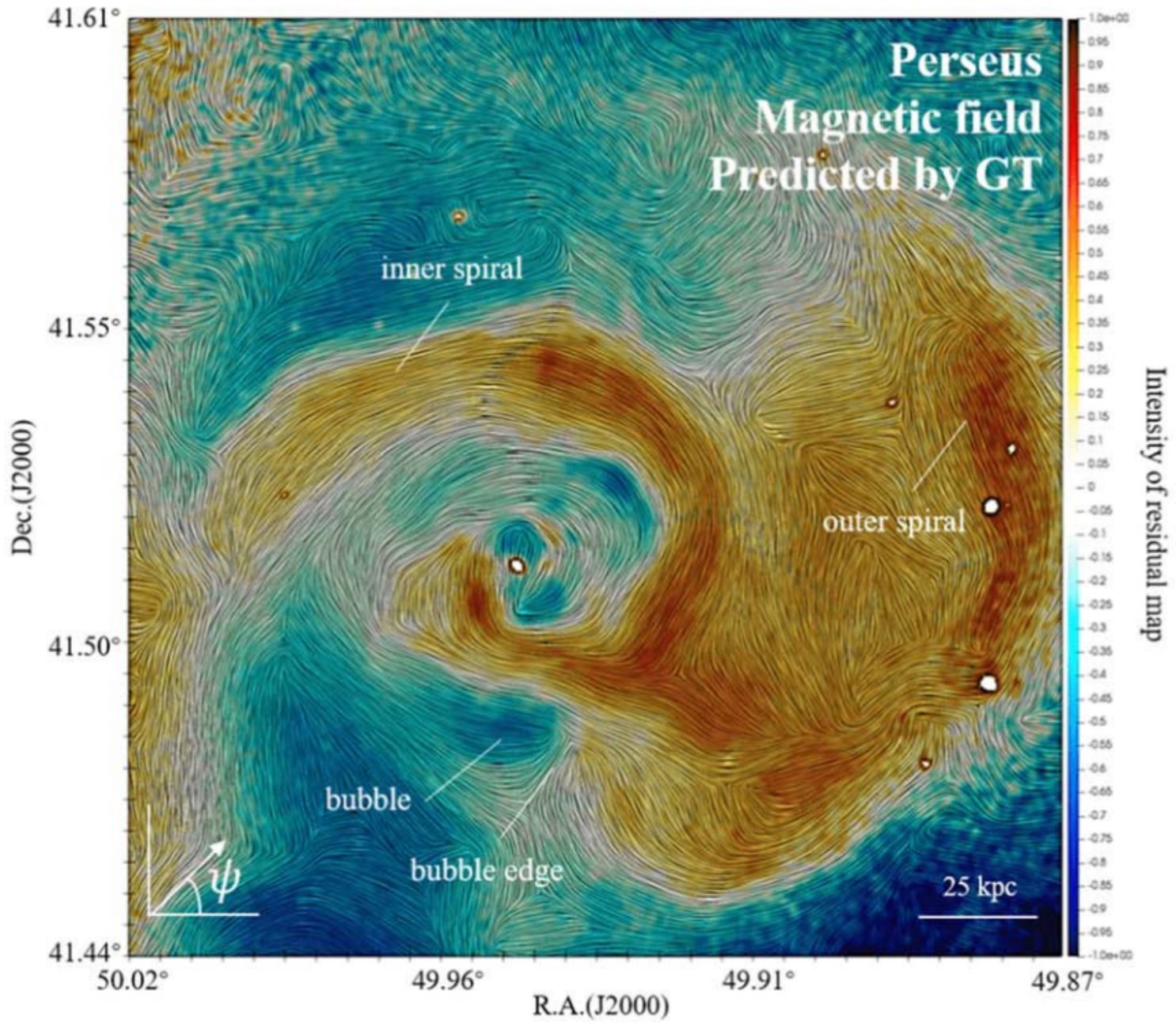}
	\caption{\label{fig:cluster} Velocity gradients obtained from publicly available $^{12}CO$ maps from ALMA provide a better resolution magnetic field maps compared to VLA.
 From Hu et al. (2022c)}
\end{figure}

The GT can provide magnetic fields in situations where all other techniques fail. For instance, in \cite{hu2019magnetic}, the POS magnetic fields were mapped with VChGs for a tenuous high-velocity Smith cloud, which is impossible with any other existing technique. Fig. \ref{fig:cluster} presents a similar case where the unique abilities of gradients allow magnetic field studies of galaxy clusters. The Chandra X-ray emission is used for mapping. For subsonic turbulence the density inhomogeneities that control X-ray emission mimic velocity fluctuations. Thus the Intensity Gradients (IGs) act similar to velocity gradients.

\section{Discussion}

This review provides a brief outlook on the modern state of the theory of MHD turbulence and its implications. The theory of MHD turbulence is an area of intensive research, and new discoveries of subtle turbulence properties are expected. However, despite the available advances, the MHD turbulence theory is a powerful tool for exploring astrophysical processes.

We have discussed a wide variety of processes radically changed by turbulence. Those include cosmic ray propagation and acceleration, star formation, and dynamo. This review does not get into depth about the particular applications. Instead, it provides a guide for researchers interested in knowing more about various astrophysical applications of MHD turbulence theory. A more thorough study is possible with the original papers and focused reviews we refer to.

In the review, we emphasize the intrinsic deep connection between the theory of MHD turbulence and the theory of turbulent reconnection. The fundamental properties of MHD turbulence, e.g., magnetic field wandering, cannot be understood without understanding turbulent magnetic reconnection. In fact, it is the turbulent reconnection that makes the description of MHD turbulence self-consistent. In magnetized turbulent fluids, turbulent reconnection is responsible for a process of {\it reconnection diffusion} that removes magnetic flux from molecular clouds and resolves the problem of catastrophic breaking of matter in circumstellar disks.   

To exemplify the power of MHD turbulence theory, we have discussed the technique of magnetic field study that utilizes the properties of MHD turbulence; particularly, the turbulent reconnection is a part a parcel of the MHD turbulent cascade. This new technique, called Gradient Technique (GT), has proven to be a powerful tool for studying magnetic fields in the Milky Way, nearby galaxies, and galaxy clusters.

\section*{Acknowledgements}
A.L. acknowledges the support of NASA ATP AAH7546.



\bibliographystyle{mnras}
\bibliography{example} 

\begin{thebibliography}{}
\makeatletter
\relax
\def\mn@urlcharsother{\let\do\@makeother \do\$\do\&\do\#\do\^\do\_\do\%\do\~}
\def\mn@doi{\begingroup\mn@urlcharsother \@ifnextchar [ {\mn@doi@}
  {\mn@doi@[]}}
\def\mn@doi@[#1]#2{\def\@tempa{#1}\ifx\@tempa\@empty \href
  {http://dx.doi.org/#2} {doi:#2}\else \href {http://dx.doi.org/#2} {#1}\fi
  \endgroup}
\def\mn@eprint#1#2{\mn@eprint@#1:#2::\@nil}
\def\mn@eprint@arXiv#1{\href {http://arxiv.org/abs/#1} {{\tt arXiv:#1}}}
\def\mn@eprint@dblp#1{\href {http://dblp.uni-trier.de/rec/bibtex/#1.xml}
  {dblp:#1}}
\def\mn@eprint@#1:#2:#3:#4\@nil{\def\@tempa {#1}\def\@tempb {#2}\def\@tempc
  {#3}\ifx \@tempc \@empty \let \@tempc \@tempb \let \@tempb \@tempa \fi \ifx
  \@tempb \@empty \def\@tempb {arXiv}\fi \@ifundefined
  {mn@eprint@\@tempb}{\@tempb:\@tempc}{\expandafter \expandafter \csname
  mn@eprint@\@tempb\endcsname \expandafter{\@tempc}}}

\bibitem[\protect\citeauthoryear{Alexander, Steven, Samuel, Jason  \&
  James}{Alexander et~al.}{2004}]{Schekochihin2004}
Alexander A.~S.,  Steven C.~C.,  Samuel F.~T.,  Jason L.~M.,   James C.~M.,
  2004, The Astrophysical Journal, 612, 276

\bibitem[\protect\citeauthoryear{Andersson, Lazarian  \&
  Vaillancourt}{Andersson et~al.}{2015}]{andersson2015interstellar}
Andersson B.,  Lazarian A.,   Vaillancourt J.~E.,  2015, Annual Review of
  Astronomy and Astrophysics, 53, 501

\bibitem[\protect\citeauthoryear{Arctowski}{Arctowski}{1940}]{Arctowski1940}
Arctowski H.,  1940, Proceedings of the National Academy of Sciences, 26, 406

\bibitem[\protect\citeauthoryear{{Armstrong}, {Rickett}  \&
  {Spangler}}{{Armstrong} et~al.}{1995}]{Armstrong_etal:1995}
{Armstrong} J.~W.,  {Rickett} B.~J.,   {Spangler} S.~R.,  1995, \mn@doi [\apj]
  {10.1086/175515}, 443, 209

\bibitem[\protect\citeauthoryear{{Balbus} \& {Hawley}}{{Balbus} \&
  {Hawley}}{1998}]{BalbusHawley:1998}
{Balbus} S.~A.,  {Hawley} J.~F.,  1998, \mn@doi [Reviews of Modern Physics]
  {10.1103/RevModPhys.70.1}, 70, 1

\bibitem[\protect\citeauthoryear{Baliunas \& Vaughan}{Baliunas \&
  Vaughan}{1985}]{Baliunas1985}
Baliunas S.~L.,  Vaughan A.~H.,  1985, Annual review of astronomy and
  astrophysics, 23, 379

\bibitem[\protect\citeauthoryear{Beck}{Beck}{2016}]{beck2016magnetic}
Beck R.,  2016, The Astronomy and Astrophysics Review, 24, 4

\bibitem[\protect\citeauthoryear{{Beresnyak}}{{Beresnyak}}{2017}]{Beresnyak:2017}
{Beresnyak} A.,  2017, \mn@doi [\apj] {10.3847/1538-4357/834/1/47}, 834, 47

\bibitem[\protect\citeauthoryear{Beresnyak}{Beresnyak}{2019}]{Beresnyak2019}
Beresnyak A.,  2019, Living Reviews in Computational Astrophysics, 5, 2

\bibitem[\protect\citeauthoryear{{Beresnyak} \& {Lazarian}}{{Beresnyak} \&
  {Lazarian}}{2019}]{BL19}
{Beresnyak} A.,  {Lazarian} A.,  2019, {Turbulence in Magnetohydrodynamics}

\bibitem[\protect\citeauthoryear{Brandenburg \& Subramanian}{Brandenburg \&
  Subramanian}{2005}]{Brandenburg2005}
Brandenburg A.,  Subramanian K.,  2005, Physics Reports, 417, 1

\bibitem[\protect\citeauthoryear{{Brunetti} \& {Jones}}{{Brunetti} \&
  {Jones}}{2014}]{Brun14}
{Brunetti} G.,  {Jones} T.~W.,  2014, \mn@doi [International Journal of Modern
  Physics D] {10.1142/S0218271814300079}, \href
  {https://ui.adsabs.harvard.edu/abs/2014IJMPD..2330007B} {23, 1430007}

\bibitem[\protect\citeauthoryear{{Burkhart}, {Stanimirovi{\'c}}, {Lazarian}  \&
  {Kowal}}{{Burkhart} et~al.}{2010}]{Burkhart_etal:2010}
{Burkhart} B.,  {Stanimirovi{\'c}} S.,  {Lazarian} A.,   {Kowal} G.,  2010,
  \mn@doi [\apj] {10.1088/0004-637X/708/2/1204}, 708, 1204

\bibitem[\protect\citeauthoryear{{Chandran}}{{Chandran}}{2005}]{Chandran:2005}
{Chandran} B.~D.~G.,  2005, \mn@doi [\apj] {10.1086/432596}, 632, 809

\bibitem[\protect\citeauthoryear{{Chepurnov} \& {Lazarian}}{{Chepurnov} \&
  {Lazarian}}{2010}]{ChepurnovLazarian:2010}
{Chepurnov} A.,  {Lazarian} A.,  2010, \mn@doi [\apj]
  {10.1088/0004-637X/710/1/853}, 710, 853

\bibitem[\protect\citeauthoryear{Chertkov, Falkovich, Kolokolov  \&
  Vergassola}{Chertkov et~al.}{1999}]{chertkov1999small}
Chertkov M.,  Falkovich G.,  Kolokolov I.,   Vergassola M.,  1999, Physical
  review letters, 83, 4065

\bibitem[\protect\citeauthoryear{{Chitta} \& {Lazarian}}{{Chitta} \&
  {Lazarian}}{2019}]{ChittaLazarian:2019}
{Chitta} L.~P.,  {Lazarian} A.,  2019, \apj, submitted

\bibitem[\protect\citeauthoryear{{Cho} \& {Lazarian}}{{Cho} \&
  {Lazarian}}{2002}]{CL02_PRL}
{Cho} J.,  {Lazarian} A.,  2002, \mn@doi [Physical Review Letters]
  {10.1103/PhysRevLett.88.245001}, \href
  {http://adsabs.harvard.edu/abs/2002PhRvL..88x5001C} {88, 245001}

\bibitem[\protect\citeauthoryear{{Cho} \& {Lazarian}}{{Cho} \&
  {Lazarian}}{2003}]{CL03}
{Cho} J.,  {Lazarian} A.,  2003, \mn@doi [\mnras]
  {10.1046/j.1365-8711.2003.06941.x}, \href
  {http://adsabs.harvard.edu/abs/2003MNRAS.345..325C} {345, 325}

\bibitem[\protect\citeauthoryear{{Cho} \& {Vishniac}}{{Cho} \&
  {Vishniac}}{2000}]{CV00}
{Cho} J.,  {Vishniac} E.~T.,  2000, \mn@doi [\apj] {10.1086/309213}, \href
  {http://adsabs.harvard.edu/abs/2000ApJ...539..273C} {539, 273}

\bibitem[\protect\citeauthoryear{{Cho}, {Lazarian}  \& {Vishniac}}{{Cho}
  et~al.}{2002a}]{CLV_incomp}
{Cho} J.,  {Lazarian} A.,   {Vishniac} E.~T.,  2002a, \mn@doi [\apj]
  {10.1086/324186}, \href {http://adsabs.harvard.edu/abs/2002ApJ...564..291C}
  {564, 291}

\bibitem[\protect\citeauthoryear{{Cho}, {Lazarian}  \& {Vishniac}}{{Cho}
  et~al.}{2002b}]{Cho_etal:2002}
{Cho} J.,  {Lazarian} A.,   {Vishniac} E.~T.,  2002b, \mn@doi [\apj]
  {10.1086/324186}, 564, 291

\bibitem[\protect\citeauthoryear{{En{\ss}lin} \& {Vogt}}{{En{\ss}lin} \&
  {Vogt}}{2006}]{EnsslinVogt:2006}
{En{\ss}lin} T.~A.,  {Vogt} C.,  2006, \mn@doi [\aap]
  {10.1051/0004-6361:20053518}, 453, 447

\bibitem[\protect\citeauthoryear{Eyink, Lazarian  \& Vishniac}{Eyink
  et~al.}{2011}]{Eyink_etal:2011}
Eyink G.~L.,  Lazarian A.,   Vishniac E.~T.,  2011, The Astrophysical Journal,
  743, 51

\bibitem[\protect\citeauthoryear{{Eyink} et~al.,}{{Eyink}
  et~al.}{2013}]{Eyink_etal:2013}
{Eyink} G.,  et~al., 2013, \mn@doi [\nat] {10.1038/nature12128}, 497, 466

\bibitem[\protect\citeauthoryear{{Farmer} \& {Goldreich}}{{Farmer} \&
  {Goldreich}}{2004}]{FG04}
{Farmer} A.~J.,  {Goldreich} P.,  2004, \mn@doi [\apj] {10.1086/382040}, \href
  {http://adsabs.harvard.edu/abs/2004ApJ...604..671F} {604, 671}

\bibitem[\protect\citeauthoryear{{Ferri{\`e}re}}{{Ferri{\`e}re}}{2001}]{Ferriere:2001}
{Ferri{\`e}re} K.~M.,  2001, \mn@doi [Reviews of Modern Physics]
  {10.1103/RevModPhys.73.1031}, 73, 1031

\bibitem[\protect\citeauthoryear{{Galli}, {Lizano}, {Shu}  \& {Allen}}{{Galli}
  et~al.}{2006}]{Galli_etal:2006}
{Galli} D.,  {Lizano} S.,  {Shu} F.~H.,   {Allen} A.,  2006, \mn@doi [\apj]
  {10.1086/505257}, \href
  {https://ui.adsabs.harvard.edu/abs/2006ApJ...647..374G} {647, 374}

\bibitem[\protect\citeauthoryear{{Galsgaard} \& {Nordlund}}{{Galsgaard} \&
  {Nordlund}}{1997}]{GalsgaardNordlund:1997}
{Galsgaard} K.,  {Nordlund} {\AA}.,  1997, \mn@doi [\jgr] {10.1029/96JA02680},
  102, 231

\bibitem[\protect\citeauthoryear{{Galtier}, {Nazarenko}, {Newell}  \&
  {Pouquet}}{{Galtier} et~al.}{2000}]{Gal00}
{Galtier} S.,  {Nazarenko} S.~V.,  {Newell} A.~C.,   {Pouquet} A.,  2000,
  \mn@doi [Journal of Plasma Physics] {10.1017/S0022377899008284}, \href
  {http://adsabs.harvard.edu/abs/2000JPlPh..63..447G} {63, 447}

\bibitem[\protect\citeauthoryear{Garland}{Garland}{1979}]{Garland1979}
Garland G.~D.,  1979, Historia Mathematica, 6, 5

\bibitem[\protect\citeauthoryear{{Gerrard} \& {Hood}}{{Gerrard} \&
  {Hood}}{2003}]{GerrardHood:2003}
{Gerrard} C.~L.,  {Hood} A.~W.,  2003, \mn@doi [\solphys]
  {10.1023/A:1024053501326}, 214, 151

\bibitem[\protect\citeauthoryear{{Giacalone} \& {Jokipii}}{{Giacalone} \&
  {Jokipii}}{1999}]{Giacalone_Jok1999}
{Giacalone} J.,  {Jokipii} J.~R.,  1999, \mn@doi [\apj] {10.1086/307452}, \href
  {http://adsabs.harvard.edu/abs/1999ApJ...520..204G} {520, 204}

\bibitem[\protect\citeauthoryear{{Goldreich} \& {Sridhar}}{{Goldreich} \&
  {Sridhar}}{1995}]{GS95}
{Goldreich} P.,  {Sridhar} S.,  1995, \mn@doi [\apj] {10.1086/175121}, \href
  {http://adsabs.harvard.edu/abs/1995ApJ...438..763G} {438, 763}

\bibitem[\protect\citeauthoryear{Gonz{\'a}lez-Casanova \&
  Lazarian}{Gonz{\'a}lez-Casanova \& Lazarian}{2017}]{gonzalez2017velocity}
Gonz{\'a}lez-Casanova D.~F.,  Lazarian A.,  2017, The Astrophysical Journal,
  835, 41

\bibitem[\protect\citeauthoryear{{Guo} \& {Oh}}{{Guo} \& {Oh}}{2008}]{Guo08}
{Guo} F.,  {Oh} S.~P.,  2008, \mn@doi [\mnras]
  {10.1111/j.1365-2966.2007.12692.x}, \href
  {https://ui.adsabs.harvard.edu/abs/2008MNRAS.384..251G} {384, 251}

\bibitem[\protect\citeauthoryear{Hale}{Hale}{2013}]{Hale2013}
Hale G.~E.,  2013, in , A Source Book in Astronomy and Astrophysics,
  1900--1975.
Harvard University Press, pp 96--105

\bibitem[\protect\citeauthoryear{Hall \& Lockwood}{Hall \&
  Lockwood}{1998}]{Hall1998}
Hall J.~C.,  Lockwood G.,  1998, The Astrophysical Journal, 493, 494

\bibitem[\protect\citeauthoryear{Ho, Yuen, Leung  \& Lazarian}{Ho
  et~al.}{2019}]{ho2019comparison}
Ho K.~W.,  Yuen K.~H.,  Leung P.~K.,   Lazarian A.,  2019, The Astrophysical
  Journal, 887, 258

\bibitem[\protect\citeauthoryear{{Holguin}, {Ruszkowski}, {Lazarian}, {Farber}
  \& {Yang}}{{Holguin} et~al.}{2019}]{Hol19}
{Holguin} F.,  {Ruszkowski} M.,  {Lazarian} A.,  {Farber} R.,   {Yang} H.
  Y.~K.,  2019, \mn@doi [\mnras] {10.1093/mnras/stz2568}, \href
  {https://ui.adsabs.harvard.edu/abs/2019MNRAS.490.1271H} {490, 1271}

\bibitem[\protect\citeauthoryear{Howard et~al.,}{Howard
  et~al.}{2020}]{Howard2020}
Howard W.~S.,  et~al., 2020, The Astrophysical Journal, 902, 115

\bibitem[\protect\citeauthoryear{Hu et~al.,}{Hu et~al.}{2019}]{hu2019magnetic}
Hu Y.,  et~al., 2019, Nature Astronomy, 3, 776

\bibitem[\protect\citeauthoryear{Hu, Lazarian, Beck  \& Xu}{Hu
  et~al.}{2022}]{hu2022role}
Hu Y.,  Lazarian A.,  Beck R.,   Xu S.,  2022, The Astrophysical Journal, 941,
  92

\bibitem[\protect\citeauthoryear{{Ipavich}}{{Ipavich}}{1975}]{Ipa75}
{Ipavich} F.~M.,  1975, \mn@doi [\apj] {10.1086/153397}, \href
  {https://ui.adsabs.harvard.edu/abs/1975ApJ...196..107I} {196, 107}

\bibitem[\protect\citeauthoryear{{Jafari} \& {Vishniac}}{{Jafari} \&
  {Vishniac}}{2018}]{JafariVishniac2018disks}
{Jafari} A.,  {Vishniac} E.~T.,  2018, \mn@doi [\apj]
  {10.3847/1538-4357/aaa75b}, 854, 2

\bibitem[\protect\citeauthoryear{{Johns-Krull}}{{Johns-Krull}}{2007}]{Johns-Krull:2007}
{Johns-Krull} C.~M.,  2007, \mn@doi [\apj] {10.1086/519017}, \href
  {https://ui.adsabs.harvard.edu/abs/2007ApJ...664..975J} {664, 975}

\bibitem[\protect\citeauthoryear{{Jokipii}}{{Jokipii}}{1966}]{Jokipii1966}
{Jokipii} J.~R.,  1966, \mn@doi [\apj] {10.1086/148912}, \href
  {http://adsabs.harvard.edu/abs/1966ApJ...146..480J} {146, 480}

\bibitem[\protect\citeauthoryear{{Jokipii}}{{Jokipii}}{1971}]{Jo71}
{Jokipii} J.~R.,  1971, \mn@doi [Reviews of Geophysics and Space Physics]
  {10.1029/RG009i001p00027}, \href
  {https://ui.adsabs.harvard.edu/abs/1971RvGSP...9...27J} {9, 27}

\bibitem[\protect\citeauthoryear{Kazantsev}{Kazantsev}{1968}]{Kazantsev1968}
Kazantsev A.,  1968, Sov. Phys. JETP, 26, 1031

\bibitem[\protect\citeauthoryear{{Kowal} \& {Lazarian}}{{Kowal} \&
  {Lazarian}}{2010}]{KowL10}
{Kowal} G.,  {Lazarian} A.,  2010, \mn@doi [\apj]
  {10.1088/0004-637X/720/1/742}, \href
  {http://adsabs.harvard.edu/abs/2010ApJ...720..742K} {720, 742}

\bibitem[\protect\citeauthoryear{{Kowal}, {Lazarian}, {Vishniac}  \&
  {Otmianowska-Mazur}}{{Kowal} et~al.}{2009}]{Kowal_etal:2009}
{Kowal} G.,  {Lazarian} A.,  {Vishniac} E.~T.,   {Otmianowska-Mazur} K.,  2009,
  \mn@doi [\apj] {10.1088/0004-637X/700/1/63}, 700, 63

\bibitem[\protect\citeauthoryear{{Kowal}, {de Gouveia Dal Pino}  \&
  {Lazarian}}{{Kowal} et~al.}{2011}]{Kowal_etal:2011}
{Kowal} G.,  {de Gouveia Dal Pino} E.~M.,   {Lazarian} A.,  2011, \mn@doi
  [\apj] {10.1088/0004-637X/735/2/102}, 735, 102

\bibitem[\protect\citeauthoryear{{Kowal}, {Lazarian}, {Vishniac}  \&
  {Otmianowska-Mazur}}{{Kowal} et~al.}{2012}]{Kowal_etal:2012a}
{Kowal} G.,  {Lazarian} A.,  {Vishniac} E.~T.,   {Otmianowska-Mazur} K.,  2012,
  \mn@doi [Nonlinear Processes in Geophysics] {10.5194/npg-19-297-2012}, \href
  {https://ui.adsabs.harvard.edu/abs/2012NPGeo..19..297K} {19, 297}

\bibitem[\protect\citeauthoryear{{Kowal}, {Falceta-Gon{\c c}alves}, {Lazarian}
  \& {Vishniac}}{{Kowal} et~al.}{2017}]{Kowal_etal:2017}
{Kowal} G.,  {Falceta-Gon{\c c}alves} D.~A.,  {Lazarian} A.,   {Vishniac}
  E.~T.,  2017, \mn@doi [\apj] {10.3847/1538-4357/aa6001}, 838, 91

\bibitem[\protect\citeauthoryear{Kowal, Falceta-Gon{\c{c}}alves, Lazarian  \&
  Vishniac}{Kowal et~al.}{2020}]{kowal2020kelvin}
Kowal G.,  Falceta-Gon{\c{c}}alves D.~A.,  Lazarian A.,   Vishniac E.~T.,
  2020, The Astrophysical Journal, 892, 50

\bibitem[\protect\citeauthoryear{Kraichnan \& Nagarajan}{Kraichnan \&
  Nagarajan}{1967}]{Kraichnan1967}
Kraichnan R.~H.,  Nagarajan S.,  1967, The Physics of Fluids, 10, 859

\bibitem[\protect\citeauthoryear{{Kulsrud} \& {Pearce}}{{Kulsrud} \&
  {Pearce}}{1969}]{Kulsrud_Pearce}
{Kulsrud} R.,  {Pearce} W.~P.,  1969, \mn@doi [\apj] {10.1086/149981}, \href
  {http://adsabs.harvard.edu/abs/1969ApJ...156..445K} {156, 445}

\bibitem[\protect\citeauthoryear{Lazarian}{Lazarian}{2005}]{lazarian2005production}
Lazarian A.,  2005, Astronomy \& Astrophysics, 441, 845

\bibitem[\protect\citeauthoryear{{Lazarian}}{{Lazarian}}{2006}]{Lazarian06}
{Lazarian} A.,  2006, \mn@doi [\apjl] {10.1086/505796}, \href
  {http://adsabs.harvard.edu/abs/2006ApJ...645L..25L} {645, L25}

\bibitem[\protect\citeauthoryear{{Lazarian}}{{Lazarian}}{2016}]{La16}
{Lazarian} A.,  2016, \mn@doi [\apj] {10.3847/1538-4357/833/2/131}, \href
  {http://adsabs.harvard.edu/abs/2016ApJ...833..131L} {833, 131}

\bibitem[\protect\citeauthoryear{{Lazarian} \& {Vishniac}}{{Lazarian} \&
  {Vishniac}}{1999}]{LV99}
{Lazarian} A.,  {Vishniac} E.~T.,  1999, \mn@doi [\apj] {10.1086/307233}, 517,
  700

\bibitem[\protect\citeauthoryear{{Lazarian} \& {Xu}}{{Lazarian} \&
  {Xu}}{2021}]{LX21}
{Lazarian} A.,  {Xu} S.,  2021, \mn@doi [\apj] {10.3847/1538-4357/ac2de9},
  \href {https://ui.adsabs.harvard.edu/abs/2021ApJ...923...53L} {923, 53}

\bibitem[\protect\citeauthoryear{{Lazarian} \& {Xu}}{{Lazarian} \&
  {Xu}}{2022}]{2022FrP....10.2799L}
{Lazarian} A.,  {Xu} S.,  2022, \mn@doi [Frontiers in Physics]
  {10.3389/fphy.2022.702799}, \href
  {https://ui.adsabs.harvard.edu/abs/2022FrP....10.2799L} {10, 702799}

\bibitem[\protect\citeauthoryear{{Lazarian} \& {Yan}}{{Lazarian} \&
  {Yan}}{2002}]{LY02}
{Lazarian} A.,  {Yan} H.,  2002, \mn@doi [\apjl] {10.1086/339675}, \href
  {http://adsabs.harvard.edu/abs/2002ApJ...566L.105L} {566, L105}

\bibitem[\protect\citeauthoryear{{Lazarian} \& {Yan}}{{Lazarian} \&
  {Yan}}{2014}]{LazarianYan:2014}
{Lazarian} A.,  {Yan} H.,  2014, \mn@doi [\apj] {10.1088/0004-637X/784/1/38},
  784, 38

\bibitem[\protect\citeauthoryear{{Lazarian} \& {Yuen}}{{Lazarian} \&
  {Yuen}}{2018a}]{LazarianYuen:2018a}
{Lazarian} A.,  {Yuen} K.~H.,  2018a, \mn@doi [\apj]
  {10.3847/1538-4357/aaa241}, \href
  {https://ui.adsabs.harvard.edu/abs/2018ApJ...853...96L} {853, 96}

\bibitem[\protect\citeauthoryear{{Lazarian} \& {Yuen}}{{Lazarian} \&
  {Yuen}}{2018b}]{LazarianYuen:2018b}
{Lazarian} A.,  {Yuen} K.~H.,  2018b, \mn@doi [\apj]
  {10.3847/1538-4357/aad3ca}, \href
  {https://ui.adsabs.harvard.edu/abs/2018ApJ...865...59L} {865, 59}

\bibitem[\protect\citeauthoryear{Lazarian, Yuen, Lee  \& Cho}{Lazarian
  et~al.}{2017}]{lazarian2017synchrotron}
Lazarian A.,  Yuen K.~H.,  Lee H.,   Cho J.,  2017, The Astrophysical Journal,
  842, 30

\bibitem[\protect\citeauthoryear{Lazarian, Zhang  \& Xu}{Lazarian
  et~al.}{2019}]{lazarian2019gamma}
Lazarian A.,  Zhang B.,   Xu S.,  2019, The Astrophysical Journal, 882, 184

\bibitem[\protect\citeauthoryear{Lazarian, Eyink, Jafari, Kowal, Li, Xu  \&
  Vishniac}{Lazarian et~al.}{2020}]{lazarian20203d}
Lazarian A.,  Eyink G.~L.,  Jafari A.,  Kowal G.,  Li H.,  Xu S.,   Vishniac
  E.~T.,  2020, Physics of Plasmas, 27, 012305

\bibitem[\protect\citeauthoryear{{Leamon}, {Smith}, {Ness}, {Matthaeus}  \&
  {Wong}}{{Leamon} et~al.}{1998}]{Leamon_etal:1998}
{Leamon} R.~J.,  {Smith} C.~W.,  {Ness} N.~F.,  {Matthaeus} W.~H.,   {Wong}
  H.~K.,  1998, \mn@doi [\jgr] {10.1029/97JA03394}, 103, 4775

\bibitem[\protect\citeauthoryear{{Lithwick} \& {Goldreich}}{{Lithwick} \&
  {Goldreich}}{2001}]{LG01}
{Lithwick} Y.,  {Goldreich} P.,  2001, \mn@doi [\apj] {10.1086/323470}, \href
  {http://adsabs.harvard.edu/abs/2001ApJ...562..279L} {562, 279}

\bibitem[\protect\citeauthoryear{{Longair}}{{Longair}}{2011}]{2011hea..book.....L}
{Longair} M.~S.,  2011, {High Energy Astrophysics}

\bibitem[\protect\citeauthoryear{{Maron} \& {Goldreich}}{{Maron} \&
  {Goldreich}}{2001}]{MG01}
{Maron} J.,  {Goldreich} P.,  2001, \mn@doi [\apj] {10.1086/321413}, \href
  {http://adsabs.harvard.edu/abs/2001ApJ...554.1175M} {554, 1175}

\bibitem[\protect\citeauthoryear{{Matthaeus}, {Goldstein}  \&
  {Roberts}}{{Matthaeus} et~al.}{1990}]{Mat90}
{Matthaeus} W.~H.,  {Goldstein} M.~L.,   {Roberts} D.~A.,  1990, \mn@doi [\jgr]
  {10.1029/JA095iA12p20673}, \href
  {http://adsabs.harvard.edu/abs/1990JGR....9520673M} {95, 20673}

\bibitem[\protect\citeauthoryear{{Matthaeus}, {Ghosh}, {Oughton}  \&
  {Roberts}}{{Matthaeus} et~al.}{1996}]{1996JGR...101.7619M}
{Matthaeus} W.~H.,  {Ghosh} S.,  {Oughton} S.,   {Roberts} D.~A.,  1996,
  \mn@doi [\jgr] {10.1029/95JA03830}, \href
  {https://ui.adsabs.harvard.edu/abs/1996JGR...101.7619M} {101, 7619}

\bibitem[\protect\citeauthoryear{Mclntosh \& Dryer}{Mclntosh \&
  Dryer}{1972}]{Mclntosh1972}
Mclntosh P.~S.,  Dryer M.,  1972, Solar activity observations and predictions.
American Institute of Aeronautics and Astronautics

\bibitem[\protect\citeauthoryear{{Mestel}}{{Mestel}}{1965a}]{Mestel:1965a}
{Mestel} L.,  1965a, \qjras, \href
  {https://ui.adsabs.harvard.edu/abs/1965QJRAS...6..161M} {6, 161}

\bibitem[\protect\citeauthoryear{{Mestel}}{{Mestel}}{1965b}]{Mestel:1965b}
{Mestel} L.,  1965b, \qjras, \href
  {https://ui.adsabs.harvard.edu/abs/1965QJRAS...6..265M} {6, 265}

\bibitem[\protect\citeauthoryear{{Mestel}}{{Mestel}}{1966}]{Mestel:1966}
{Mestel} L.,  1966, \mn@doi [\mnras] {10.1093/mnras/133.2.265}, \href
  {https://ui.adsabs.harvard.edu/abs/1966MNRAS.133..265M} {133, 265}

\bibitem[\protect\citeauthoryear{{Mestel} \& {Spitzer}}{{Mestel} \&
  {Spitzer}}{1956}]{MestelSpitzer:1956}
{Mestel} L.,  {Spitzer} L. J.,  1956, \mn@doi [\mnras]
  {10.1093/mnras/116.5.503}, \href
  {https://ui.adsabs.harvard.edu/abs/1956MNRAS.116..503M} {116, 503}

\bibitem[\protect\citeauthoryear{Moffatt}{Moffatt}{1978}]{Moffatt1978}
Moffatt H.~K.,  1978, Cambridge University Press, Cambridge, London, New York,
  Melbourne, 2, 5

\bibitem[\protect\citeauthoryear{Mouschovias}{Mouschovias}{1991}]{mouschovias1991magnetic}
Mouschovias T.~C.,  1991, Astrophysical Journal, Part 1 (ISSN 0004-637X), vol.
  373, May 20, 1991, p. 169-186. NSF-supported research., 373, 169

\bibitem[\protect\citeauthoryear{{Mouschovias}, {Tassis}  \&
  {Kunz}}{{Mouschovias} et~al.}{2006}]{Mouschovias_etal:2006}
{Mouschovias} T.~C.,  {Tassis} K.,   {Kunz} M.~W.,  2006, \mn@doi [\apj]
  {10.1086/500125}, \href
  {https://ui.adsabs.harvard.edu/abs/2006ApJ...646.1043M} {646, 1043}

\bibitem[\protect\citeauthoryear{Nakano, Nishi  \& Umebayashi}{Nakano
  et~al.}{2002}]{nakano2002mechanism}
Nakano T.,  Nishi R.,   Umebayashi T.,  2002, The Astrophysical Journal, 573,
  199

\bibitem[\protect\citeauthoryear{{Norman} \& {Ferrara}}{{Norman} \&
  {Ferrara}}{1996}]{NormanFerrara:1996}
{Norman} C.~A.,  {Ferrara} A.,  1996, \mn@doi [\apj] {10.1086/177603}, 467, 280

\bibitem[\protect\citeauthoryear{{Oishi}, {Mac Low}, {Collins}  \&
  {Tamura}}{{Oishi} et~al.}{2015}]{Oishi_etal:2015}
{Oishi} J.~S.,  {Mac Low} M.-M.,  {Collins} D.~C.,   {Tamura} M.,  2015,
  preprint

\bibitem[\protect\citeauthoryear{Parker}{Parker}{1955}]{Parker1955}
Parker E.~N.,  1955, The Astrophysical Journal, 122, 293

\bibitem[\protect\citeauthoryear{{Parker}}{{Parker}}{1965}]{Par65}
{Parker} E.~N.,  1965, \mn@doi [\planss] {10.1016/0032-0633(65)90131-5}, \href
  {https://ui.adsabs.harvard.edu/abs/1965P&SS...13....9P} {13, 9}

\bibitem[\protect\citeauthoryear{Rincon}{Rincon}{2019}]{Rincon2019}
Rincon F.,  2019, Journal of Plasma Physics, 85, 205850401

\bibitem[\protect\citeauthoryear{{Santos-Lima}, {Lazarian}, {de Gouveia Dal
  Pino}  \& {Cho}}{{Santos-Lima} et~al.}{2010}]{Santos-Lima_etal:2010}
{Santos-Lima} R.,  {Lazarian} A.,  {de Gouveia Dal Pino} E.~M.,   {Cho} J.,
  2010, \mn@doi [\apj] {10.1088/0004-637X/714/1/442}, \href
  {https://ui.adsabs.harvard.edu/abs/2010ApJ...714..442S} {714, 442}

\bibitem[\protect\citeauthoryear{Santos-Lima, Guerrero, de Gouveia Dal~Pino  \&
  Lazarian}{Santos-Lima et~al.}{2021}]{santos2021diffusion}
Santos-Lima R.,  Guerrero G.,  de Gouveia Dal~Pino E.,   Lazarian A.,  2021,
  Monthly Notices of the Royal Astronomical Society, 503, 1290

\bibitem[\protect\citeauthoryear{{Schlickeiser}}{{Schlickeiser}}{2002a}]{Schlickeiser02}
{Schlickeiser} R.,  2002a, {Cosmic Ray Astrophysics}

\bibitem[\protect\citeauthoryear{{Schlickeiser}}{{Schlickeiser}}{2002b}]{Schlickeiser2002}
{Schlickeiser} R.,  2002b, {Cosmic Ray Astrophysics}.
Springer, Berlin

\bibitem[\protect\citeauthoryear{{Schlickeiser} \& {Miller}}{{Schlickeiser} \&
  {Miller}}{1998}]{SchlickeiserMiller}
{Schlickeiser} R.,  {Miller} J.~A.,  1998, \mn@doi [\apj] {10.1086/305023},
  \href {http://adsabs.harvard.edu/abs/1998ApJ...492..352S} {492, 352}

\bibitem[\protect\citeauthoryear{{Shu}, {Adams}  \& {Lizano}}{{Shu}
  et~al.}{1987}]{Shu_etal:1987}
{Shu} F.~H.,  {Adams} F.~C.,   {Lizano} S.,  1987, \mn@doi [\araa]
  {10.1146/annurev.aa.25.090187.000323}, \href
  {https://ui.adsabs.harvard.edu/abs/1987ARA&A..25...23S} {25, 23}

\bibitem[\protect\citeauthoryear{{Shu}, {Li}  \& {Allen}}{{Shu}
  et~al.}{2004}]{Shu_etal:2004}
{Shu} F.~H.,  {Li} Z.-Y.,   {Allen} A.,  2004, \mn@doi [\apj] {10.1086/380602},
  \href {https://ui.adsabs.harvard.edu/abs/2004ApJ...601..930S} {601, 930}

\bibitem[\protect\citeauthoryear{{Singer}, {Heckman}  \& {Hirman}}{{Singer}
  et~al.}{2001}]{SinH01}
{Singer} H.~J.,  {Heckman} G.~R.,   {Hirman} J.~W.,  2001, \mn@doi [Washington
  DC American Geophysical Union Geophysical Monograph Series]
  {10.1029/GM125p0023}, \href
  {https://ui.adsabs.harvard.edu/abs/2001GMS...125...23S} {125, 23}

\bibitem[\protect\citeauthoryear{{Subramanian}, {Shukurov}  \&
  {Haugen}}{{Subramanian} et~al.}{2006}]{Subramanian_etal:2006}
{Subramanian} K.,  {Shukurov} A.,   {Haugen} N.~E.~L.,  2006, \mn@doi [\mnras]
  {10.1111/j.1365-2966.2006.09918.x}, 366, 1437

\bibitem[\protect\citeauthoryear{{Sych}, {Karlick{\'y}}, {Altyntsev},
  {Dud\'{\i}k}  \& {Kashapova}}{{Sych} et~al.}{2015}]{Sych_etal:2015}
{Sych} R.,  {Karlick{\'y}} M.,  {Altyntsev} A.,  {Dud\'{\i}k} J.,   {Kashapova}
  L.,  2015, \mn@doi [\aap] {10.1051/0004-6361/201424834}, 577, A43

\bibitem[\protect\citeauthoryear{Tobias}{Tobias}{2021}]{Tobias2021}
Tobias S.,  2021, Journal of fluid mechanics, 912, P1

\bibitem[\protect\citeauthoryear{Xu}{Xu}{2019}]{Xu2019}
Xu S.,  2019, Study on Magnetohydrodynamic Turbulence and Its Astrophysical
  Applications.
Springer

\bibitem[\protect\citeauthoryear{Xu \& Lazarian}{Xu \&
  Lazarian}{2016}]{xu2016turbulent}
Xu S.,  Lazarian A.,  2016, The Astrophysical Journal, 833, 215

\bibitem[\protect\citeauthoryear{{Xu} \& {Lazarian}}{{Xu} \&
  {Lazarian}}{2018}]{XLb18}
{Xu} S.,  {Lazarian} A.,  2018, \mn@doi [\apj] {10.3847/1538-4357/aae840},
  \href {https://ui.adsabs.harvard.edu/abs/2018ApJ...868...36X} {868, 36}

\bibitem[\protect\citeauthoryear{Xu \& Lazarian}{Xu \& Lazarian}{2021}]{Xu2021}
Xu S.,  Lazarian A.,  2021, Reviews of Modern Plasma Physics, 5, 2

\bibitem[\protect\citeauthoryear{{Xu} \& {Lazarian}}{{Xu} \&
  {Lazarian}}{2022}]{2022ApJ...925...48X}
{Xu} S.,  {Lazarian} A.,  2022, \mn@doi [\apj] {10.3847/1538-4357/ac3824},
  \href {https://ui.adsabs.harvard.edu/abs/2022ApJ...925...48X} {925, 48}

\bibitem[\protect\citeauthoryear{Xu \& Zhang}{Xu \&
  Zhang}{2017}]{xu2017scatter}
Xu S.,  Zhang B.,  2017, The Astrophysical Journal, 835, 2

\bibitem[\protect\citeauthoryear{{Yan} \& {Lazarian}}{{Yan} \&
  {Lazarian}}{2002}]{YL02}
{Yan} H.,  {Lazarian} A.,  2002, \mn@doi [Physical Review Letters]
  {10.1103/PhysRevLett.89.281102}, \href
  {http://adsabs.harvard.edu/abs/2002PhRvL..89B1102Y} {89, B1102+}

\bibitem[\protect\citeauthoryear{Yuen \& Lazarian}{Yuen \&
  Lazarian}{2017}]{yuen2017tracing}
Yuen K.~H.,  Lazarian A.,  2017, The Astrophysical Journal Letters, 837, L24

\bibitem[\protect\citeauthoryear{{Zuckerman} \& {Evans}}{{Zuckerman} \&
  {Evans}}{1974}]{ZuckermanEvans:1974}
{Zuckerman} B.,  {Evans} N.~J. I.,  1974, \mn@doi [\apjl] {10.1086/181613},
  \href {https://ui.adsabs.harvard.edu/abs/1974ApJ...192L.149Z} {192, L149}

\makeatother
\end{thebibliography}




\bsp	
\label{lastpage}
\end{document}